\documentclass[aps,showpacs,amsmath,amssymb,superscriptaddress,pra,twocolumn,longbibliography]{revtex4-1}

\usepackage{graphicx}
\usepackage{dcolumn}
\usepackage{bm}
\usepackage[latin1]{inputenc}
\usepackage{epstopdf}
\usepackage{amsmath}
\usepackage{amssymb}
\usepackage{color}
\usepackage{latexsym}
\usepackage{mathrsfs}

\newcommand{\e}{\mathrm{e}}

\begin{document}

%\preprint{}

\title{Mirror-assisted coherent backscattering from the Mollow sidebands}

%\author{Pipol}
%\affiliation{Somewhere}
\author{N. Piovella}
\affiliation{Dipartimento di Fisica, Universit\`{a} degli Studi di Milano, Via Celoria 16, I-20133 Milano, Italy}
%\author{P.H. Moriya}
%\affiliation{Instituto de F\'{\i}sica de S\~{a}o Carlos, Universidade de S\~{a}o Paulo, C.P. 369, 13560-970 S\~{a}o Carlos, SP, Brazil}
%\author{R.F. Shiozaki}
\author{R. Celistrino Teixeira}
\affiliation{Departamento de F\'{\i}sica, Universidade Federal de S\~{a}o Carlos, Rod. Washington Lu\'{\i}s, km 235 - SP-310, 13565-905 S\~{a}o Carlos, SP, Brazil}
%\author{C.E. M\'aximo}
%\affiliation{Instituto de F\'{\i}sica de S\~{a}o Carlos, Universidade de S\~{a}o Paulo, C.P. 369, 13560-970 S\~{a}o Carlos, SP, Brazil}
\author{R. Kaiser}
\affiliation{Universit\'e C\^ote d'Azur, CNRS, INPHYNI, France}
\author{Ph.W. Courteille}
\affiliation{Instituto de F\'{\i}sica de S\~{a}o Carlos, Universidade de S\~{a}o Paulo, C.P. 369, 13560-970 S\~{a}o Carlos, SP, Brazil}
\author{R. Bachelard}
\affiliation{Departamento de F\'{\i}sica, Universidade Federal de S\~{a}o Carlos, Rod. Washington Lu\'{\i}s, km 235 - SP-310, 13565-905 S\~{a}o Carlos, SP, Brazil}
\email{bachelard.romain@gmail.com}

\begin{abstract}
%We study the coherent resonant backscattering of light by large clouds of saturated atoms. 
In front of a mirror, the radiation of weakly driven large disordered clouds presents an interference fringe in the backward direction, on top of an incoherent background. Although strongly driven atoms usually present little coherent scattering, we here show that the mirror-assisted version can produce high contrast fringes, for arbitrarily high saturation parameters. The contrast of the fringes oscillates with the Rabi frequency of the atomic transition and the distance between the mirror and the atoms, due to the coherent interference between the carrier and the Mollow sidebands of the saturated resonant fluorescence spectrum emitted by the atoms. The setup thus represents a powerful platform to study the spectral properties of ensembles of correlated scatterers.
\end{abstract}

\maketitle

\section{Introduction}

Scattering techniques are a powerful tool to detect order in matter. When the wavelength of light becomes commensurate with a length scale of an ordered structure of scatterers, constructive interferences lead to a strongly directional emission, a phenomenon known as Bragg scattering, in clear analogy with Young's double slit experiment~\cite{Wolf2016}. Bragg scattering techniques have turned in a fundamental tool in crystallography and many other fields.

In disordered systems intuitively one rather expects an incoherent (destructive) sum of waves, yet several phenomena based on constructive interference have been identified, as for example the Coherent BackScattering of light (CBS). CBS relies on the constructive interference of two reciprocal paths, and leaves a clear signature of fringes in the backward scattering which presents an intensity higher than the radiation background. Observed for light, acoustic, seismic and matter waves, it relies on the symmetry between (time-reversed) reciprocal paths of multiple scattering~\cite{Kuga1984, Albada1985,Wolf1985,Yoo1990, Mishchenko1993, Bayer1993, Wiersma1995, Tourin1997,Larose2004, Jendrzejewski2012}.

Cold atoms have been a popular medium to study CBS of light, due to the high level of control of the light-matter coupling which can be achieved, and to the relative absence of decoherence mechanisms and inhomogeneous broadening. Still, several effects can affect the symmetry between the reciprocal paths for a cold atomic sample, which in turn reduces the contrast of the CBS cone. These can be the presence of an internal structure for the atoms~\cite{Labeyrie1999,Jonckheere2000,Bidel2002,Kupriyanov2003}, the saturation of the atoms as a which--path information becomes available through the inelastically scattered waves~\cite{Chaneliere2004,Kupriyanov2006}, or other mechanisms~\cite{Labeyrie2008}. However a quantum-mechanical treatment of even only double-scattering phenomena is a daunting task, so the proper tools to describe accurately the CBS in highly saturated atoms are still missing~\cite{Wellens2004,Totsuka1999, Balik2005,Shatokhin2005}.

We here address the problem of interferences in saturated disordered atomic systems in a somewhat simpler setup, when an optically dilute cloud is put in front of a mirror. Excited by an incident laser light beam and by its reflection on the mirror, the atoms and their mirror images generate a fringe pattern that resists to the disorder-averaging of large clouds. This mirror-assisted coherent backscattering process, hereafter called mCBS, has been studied initially in the linear optics regime~\cite{Greffet1991, Labeyrie2000}. In the case of saturated atoms, the contrast was shown to reduce as the saturation of the atomic transition increases, yet at a much lower rate than for CBS~\cite{Moriya2016}. Indeed, mCBS relies on single scattering for strongly correlated atoms, i.e., the interference of the radiation of an atom and its mirror image, rather than scattering by two or more atoms as in the case of CBS.

An important difference between the mCBS set-up and those relying on multiple scattering within the cloud is the travel time necessary to reach the mirror. Consequently, the different spectral components of the light scattered inelastically by saturated atoms spread out in phase, and one could naively expect that this would weaken the fringes' contrast (an effect which was absent in ~\cite{Moriya2016}, since for that experiment the distance to the mirror was not enough to probe the frequency broadening of the fluorescence light). Studying the quantum properties of the mCBS set-up, we here show that, contrarily to this naive expectation, the specific structure of the Mollow fluorescence spectrum allows for a high contrast even in the strongly saturated limit, provided the optical path to the mirror is well chosen. Within the right optical path, the sidebands of the Mollow triplet~\cite{Mollow1969} can be made to interfere constructively, even if averaged over the many atoms of a disordered cloud.

The Mollow spectrum of highly saturated scatterers has been first measured for an atomic beam~\cite{Schuda1974, Wu1975,Stalgies1996}, and since then for several other highly driven physical systems such as single molecules in a solid substrate~\cite{Wrigge2007}, quantum dots~\cite{Muller2007, Vamivakas2009, Flagg2009}, vacancy centers in diamond~\cite{Zhou2017}. The spectrum of a strongly driven system is proportional to the Fourier transform of the first order optical coherence, and gives valuable information about the coherent internal dynamics of the emitter and the environment that causes its decoherence~\cite{Saiko2014}: Emitters in squeezed vacuum show Mollow peaks with modified width and relative weight~\cite{Toyli2016}, and emitters coupled to cavities~\cite{Kim2014} or to other emitters~\cite{Ott2013,Pucci2017} can present high asymmetries between the two Mollow sidebands. In our setup, the dynamics of the first order optical coherence is mapped 
onto the dependence of the
contrast of spatial interference fringes on the mirror distance.
to the contrast of spatial interference fringes on the mirror distance. 
%The signal of these fringes is proportional to the number of scatterers in the system, and has thus no fundamental limitation. 
Moreover, we show here that the fringes can be obtained for arbitrarily high Rabi frequencies, which turns the mirror-assisted configuration into a powerful platform to study the quantum optics properties of strongly-driven scatterers.

This paper is organized as follows: In Sec.~\ref{sec:2}, we derive the spatial mCBS fluorescence profile for a single atom. In Sec.~\ref{sec:3}, we extend these results to a disorder-averaged cloud of scatterers. In Sec.~\ref{sec:4}, we analyze in detail the dependence of the atomic fluorescence spectrum on the observation direction and on the atomic position in the cloud, to better understand the survival of the contrast after disorder-averaging. In Sec.~\ref{sec:5}, we state our main conclusions and perspectives.

\section{Radiation from a single atom and its mirror image\label{sec:2}}

\begin{figure}
\centering
\includegraphics[width=1\linewidth]{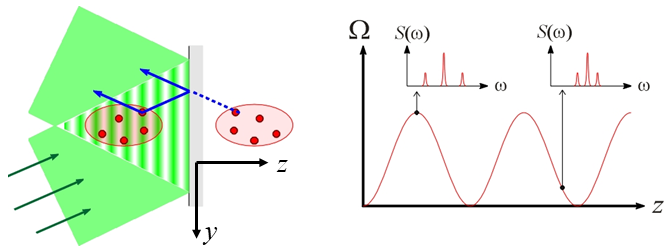}
\caption{\label{fig:Scheme} Left: Scheme of the experiment, where an incident beam is reflected on a mirror, thus creating a stationary wave. The emission is collected from the atoms and their mirror image. Right: Rabi frequency along the cloud, due to the stationary wave. The two insets are examples of inelastic spectra radiated by atoms at different positions of the stationary wave.}
\end{figure}

%\subsection{A single atom and its mirror image}
Let us first consider a single two-level atom at position $\mathbf{r} = (x,y,z)$, placed in front of a mirror which lies at the plane $z = 0$ (see Fig.~\ref{fig:Scheme}). When illuminated by an incident laser of Rabi frequency $\Omega_0$, the atom gets excited by both the laser and its reflection at the mirror, and its radiation sums up with that of its mirror image. The wavevector of the incident light, described as a plane wave, reads $(0,-k\sin\theta_0,k\cos\theta_0)$, with $\theta_0\ll 1$ the incidence angle and $k$ the laser light wavenumber. 

The superposition of both incoming and reflected laser beams create a standing wave along $z$ and a propagating wave along $y$, and the Rabi frequency $\Omega (\mathbf{r})$ seen by the atom is given by
\begin{equation}
\Omega\left(\mathbf{r}\right)=2\Omega_{0}\cos (k z\cos\theta_0)e^{-i ky\sin\theta_0}.
\label{eq:RabiFreq}
\end{equation}
with $\Omega_{0}$ the homogeneous Rabi frequency of the incident plane wave. In the semiclassical limit, the atomic dynamics is described by the well-known Bloch equations. Calling $\hat\sigma$, $\hat\sigma^{\dagger}$ and $\hat\sigma^z$ the atomic operators, these equations read~\cite{Scully1997}\begin{eqnarray}
\frac{d\hat\sigma }{dt}&=&\left(i\Delta-\frac{\Gamma}{2}\right) \hat\sigma +i\frac{\Omega(\mathbf{r})}{2}\hat\sigma^{z} ,
\\ \frac{d\hat\sigma^{z}}{dt}&=&i\left[\Omega^{*}\left(\mathbf{r}\right) \hat\sigma -\Omega\left(\mathbf{r}\right) \hat\sigma^{\dagger} \right]-\Gamma\left(\hat\sigma^{z}+1\right),
\end{eqnarray}
with the commutation relations $\left[\hat{\sigma},\hat{\sigma}^{z}\right]=2\hat{\sigma}$ and $\left[\hat{\sigma}^{\dagger},\hat{\sigma}\right]=\hat{\sigma}^{z}$. In the far-field limit, the field emitted by an atom at $\mathbf{r}$ that reaches a detector at a point $\mathbf{R}$ and at a time $t$ is given by
%\begin{eqnarray}
%\hat{\mathbf{E}}_s\left(\mathbf{R},t\right) =&&\frac{d k^{3}}{4\pi\epsilon_{0}}\left[\hat{\mathbf{x}}-\hat{\mathbf{\mathbf{n}}}\left(\hat{\mathbf{x}}\cdot\hat{\mathbf{n}}\right)\right]\hat\sigma\left(t-\frac{|\mathbf{R}-\mathbf{r}|}{c}\right)\nonumber\\
%&\times & \frac{e^{ik|\mathbf{R}-\mathbf{r}|-i\omega t}}{k|\mathbf{R}-\mathbf{r}|},
%\end{eqnarray}
\begin{equation}
\hat E_s=\frac{d k^{2}}{4\pi\epsilon_{0} R}\,\e^{i(kR-\omega t)}\, \hat\sigma\left(t - \frac{R}{c} + \frac{\hat{\mathbf{n}}\cdot\mathbf{r}}{c}\right)\e^{-i\mathbf{k}\cdot\mathbf{r}}
\end{equation}
where $\hat{\mathbf{n}} = \mathbf{k}/k \cong \mathbf{R}/R$ is the unitary vector pointing in the $\mathbf{R}-\mathbf{r}$ direction ($\mathbf{k} = k (\sin \theta \cos \varphi, \sin \theta \sin \varphi, -\cos \theta)$), $d$ is the electric dipole transition matrix element, $\epsilon_0$ the vacuum permittivity and $c$ the speed of light. 

%In the far field limit, we write $\hat{\mathbf{n}} = \mathbf{k}/k \cong \mathbf{R}/R$,
%If $\theta\ll 1$, the field $\hat{\mathbf{E}}_{s}$ is composed mainly of the $x$ component, hereafter called $\hat E_s$. Within these conditions, the field amplitude scattered by a single atom and observed in the direction $\hat{\mathbf{n}}$ writes 
%\begin{equation}
%\hat E_s=\frac{d k^{2}}{4\pi\epsilon_{0} R}\,\e^{i(kR-\omega t)}\, \hat\sigma\left(t - \frac{R}{c} + \frac{\hat{\mathbf{n}}\cdot\mathbf{r}}{c}\right)\e^{-i\mathbf{k}\cdot\mathbf{r}}
%\end{equation}

%, up to a prefactor and a phasor $\exp[i(kR-\omega t)]$.

%The distance between the mirror and the atoms needs to be reasonably small in order to produce fringes which can be resolved, but small distances are easily achieved by imaging a real mirror (RM), using a couple of lenses, into a virtual mirror (VM) at a few millimeters from the cloud. This set-up, described in Fig.XXX, allows for an easy tuning of the VM position, but also makes detectable fringes compatible with huge optical paths until the VM. 

Now, in the presence of the mirror, the radiation detected at a point $\mathbf{R}$ is composed of the radiation emitted in this direction, plus the radiation reflected in this direction by the mirror, which was first emitted in a direction $\mathbf{k}'=k\hat{\mathbf{n}}'=k (\sin\theta \cos \varphi,\sin\theta \sin \varphi,\cos\theta)$. Summing both contributions and considering the steady-state situation, we obtain 
\begin{equation}
\hat E_s(\mathbf{k},t)= \frac{d k^{2}}{4\pi\epsilon_{0} R} \hat\sigma\left(t+\frac{\hat{\mathbf{n}}\cdot\mathbf{r}}{c}\right)\e^{-i\mathbf{k}\cdot\mathbf{r}}+\hat\sigma\left(t+\frac{\hat{\mathbf{n}}'\cdot\mathbf{r}}{c}\right)\e^{-i\mathbf{k}'\cdot\mathbf{r}},\label{eq:Et}
\end{equation}
up to a phasor $\exp[i(kR-\omega t)]$ which is independent from the atom position. The intensity of the light scattered by the atom is now calculated as
\begin{multline}
I(\mathbf{k},t) = \frac{\epsilon_0 c}{2}\langle\hat E_s^\dagger (\mathbf{k},t) \hat E_s(\mathbf{k},t) \rangle \\
= I_a\left[2\langle\hat\sigma^\dagger(t)\hat\sigma(t)\rangle +2\Re\big[\langle\hat\sigma^\dagger(t)\hat\sigma(t+\tau_c)\rangle \e^{-2ikz\cos\theta }\big]\right].\label{eq:I1atom}
\end{multline}
where $I_a = d^2 k^4 c / 32 \pi^2 \epsilon_0 R^2$ and $\tau_c = (\hat{\mathbf{n}}'-\hat{\mathbf{n}})\cdot\mathbf{r}/c = 2z \cos \theta/c$ is the path difference in time units between the two contributions to the scattered light. 
One now sees that the distance between the mirror and the atom is responsible for the appearance of a two-time correlator, as the light scattered in the observation direction $\mathbf{k}$ interferes with that emitted into the direction of the mirror $\mathbf{k}'$. 
%For distances between the atom and the VM larger than a few wavelength, multiple scattering can be ignored. 
The stationary dynamics of a single atom driven by a resonant field with Rabi frequency $\Omega$ and at resonance ($\Delta=0$) is given, in the stationary regime ($t\to\infty$), by
\begin{widetext}
\begin{eqnarray}
&& \langle\sigma(t)\rangle=-\frac{i}{1+s}\frac{\Omega}{\Gamma}, \qquad \langle\sigma^+(t)\sigma(t)\rangle=\frac{s}{2(1+s)},
\\ && \left\langle\hat\sigma^\dagger\left(t\right)\hat\sigma\left(t+\tau\right)\right\rangle = \frac{s}{4(1+s)}\Bigg[\frac{2}{1+s}+\e^{-\Gamma \tau/2}
+\frac{s-1}{s+1}\cos(\Omega_M\tau)\e^{-3\Gamma\tau/4}
+\frac{\Gamma}{4\Omega_M}\frac{5s-1}{s+1}\sin(\Omega_M\tau)\e^{-3\Gamma\tau/4}\Bigg],\label{eq:2tc}
\end{eqnarray}
\end{widetext}
where we introduced the saturation parameter at resonance $s=2|\Omega(\mathbf{r})|^2/\Gamma^2$ and the Mollow frequency $\Omega_M(\mathbf{r})=\sqrt{\Omega(\mathbf{r})^2-\Gamma^2/16}$.  The fluorescence spectrum of a saturated single atom, which is given by the Fourier transform of correlator \eqref{eq:2tc}, is characterized by the emergence of sidebands, also known as the Mollow triplet~\cite{Mollow1969}. Their width is comparable to the transition linewidth, and their separation to the carrier is equal to $\Omega_M$. One then obtains from Eq.\eqref{eq:I1atom}:
\begin{widetext}
\begin{equation}
\frac{I(\mathbf{k},t)}{I_a} = \frac{s}{1+s}+\frac{s}{2(1+s)}\Bigg[\frac{2}{1+s}+\e^{-\Gamma \tau_c/2}
+\frac{s-1}{s+1}\cos(\Omega_M\tau_c)\e^{-3\Gamma\tau_c/4}
+\frac{\Gamma}{4\Omega_M}\frac{5s-1}{s+1}\sin(\Omega_M\tau_c)\e^{-3\Gamma\tau_c/4}\Bigg]\cos(2kz\cos\theta).\label{eq:I1ate}
\end{equation}
\end{widetext}

\begin{figure}
\centering
\includegraphics[width=1\linewidth]{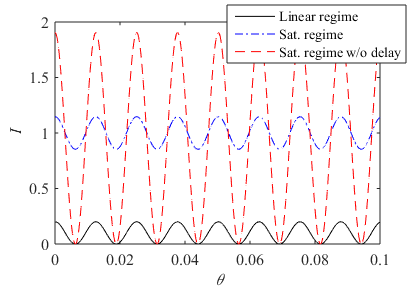}
\caption{\label{fig:SingleAtom} Angular fringe pattern from a single atom in front of a mirror  in the linear regime ($s=0.1$, plain black curve), in the saturated regime ($s=20$) in presence of a delay time ($\Omega\tau_c=3\pi/4$, dash-dotted curve) and without delay ($\tau_c=0$, dashed curve). While in the linear regime single-atom fringes always exhibit full contrast (see Eq.\eqref{low_s}), the contrast of a saturated atom will depend on the delay time (see Eq.\eqref{strong_s2}). Simulations realized for an atom at $kz=250$ with $s=0.2$ (Linear regime), $s=20$ and $\Omega_M\tau_c=3\pi/4$ (Saturated regime) and $s=20$ and $\Omega_M\tau_c=0$ (Saturated regime without delay).}
\end{figure}

Let us first discuss the low saturation case ($s\ll 1$), i.e., the linear optics regime. In this case, $\Omega_M \approx i \Gamma/4$, and the intensity can be approximated by
\begin{eqnarray}
\frac{I(\mathbf{k},t)}{I_a}&=&s[1+\cos(2kz\cos\theta)]\nonumber\\
&=&2s_0\cos^2(kz\cos\theta_0)\cos^2(kz\cos\theta),\label{low_s}
\end{eqnarray}
with $s_0=8\Omega_0^2/\Gamma^2$ the saturation parameter at the peak of the standing wave. In the weak field limit the intensity does not depend on the delay $\tau_c$ since the scattered light is emitted elastically, i.e., at the same frequency as the incident field. In this regime, a single atom in front of the mirror will exhibit an angular interference pattern with full contrast $C=(I_\text{max}-I_\text{min})/I_\text{background}=2$, where $I_\text{background}$ corresponds to the average intensity for the case of a single atom (see Fig.\ref{fig:SingleAtom}), and angular period $\pi/kz\theta_0$ around the small angle $\theta_0$. 
%Remark that it can be convenient to adjust this period independently on $\tau_c$ by imaging the mirror at an arbitrary distance from the atom~\cite{Ciaramella1993,Labeyrie2014}, as was done in a previous experiment~\cite{Moriya2016}.

We then turn to the high saturation regime ($s\gg1$), first assuming that the delay time is small compared to the transition lifetime $\Gamma^{-1}$ ($\Gamma\tau_c\ll 1$) so there is no dispersion within a single peak of the Mollow triplet, whereas it can be significant between different peaks. Eq.\eqref{eq:I1ate} then simplifies into
%\begin{eqnarray}
%I(\mathbf{k},t)&=&1+\frac{1}{2}e^{-\Gamma\tau_c/2}\nonumber\\
%&\times &
%\left[1+e^{-\Gamma\tau_c/4}\cos(\Omega_M\tau_c)\right]\cos(2kz\cos\theta)\label{strong_s}
%\end{eqnarray}
\begin{equation}
\frac{I(\mathbf{k},t)}{I_a}=1+\frac{1}{2}\left[1+\cos(\Omega_M\tau_c)\right]\cos(2kz\cos\theta).\label{strong_s2}
\end{equation}

The single-atom contrast is here given by $C=1+\cos(\Omega_M\tau_c)$, so it oscillates as the delay time $\tau_c$ or Rabi frequency $\Omega_M$ is tuned. In particular, for $\Omega_M\tau_c=0\mod (2\pi)$ the full contrast is recovered, whereas for $\Omega_M\tau_c=\pi\mod (2\pi)$, a pattern without fringes will be observed (see examples in Fig.\ref{fig:SingleAtom}). In the former case, the difference in optical path for each Mollow sideband is the same as for the central peak, so they interfere constructively and destructively at the same angles, and altogether have the same amplitude as the background. The same situation is encountered in the linear optics regime. In the latter case ($\Omega_M\tau_c=\pi\mod (2\pi)$), due to opposite interferences at each angle, the contribution of the Mollow sidebands cancels the one of the central peak, so only the background radiation is observed.

\begin{figure}
\centering
\includegraphics[width=1\linewidth]{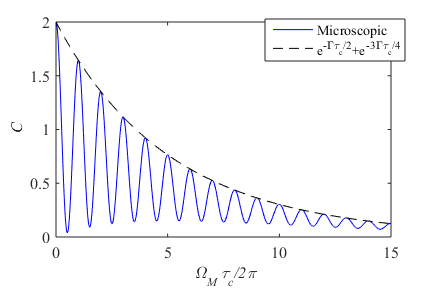}
\caption{\label{fig:SingleAtomContrast} Fringes contrast for a single atom as a function of the decoherence time $\tau_c$. Note that a plot of Eq.\eqref{eq:C1Om} overlaps extremely well with the exact contrast obtained from Eq.\eqref{eq:I1ate}, so it is not represented here. Simulations realized for $\Omega_0=10\Gamma$ with the atom at the crest of the standing wave intensity, and a laser incidence angle of $\theta_0=1^\circ$.}
\end{figure}

The growing distance between the atom and its mirror image will thus present successive drops and revivals of the contrast due to the constructive/negative interference between the Mollow sidebands and the central peak. These revivals will be eventually attenuated by the loss of coherence between the photons emitted by the atom and its mirror image as the time difference between their emission becomes of the order of $\Gamma^{-1}$ (i.e., a single peak of the Mollow triplet presents dispersion over the travel until the mirror and back). Fig.~\ref{fig:SingleAtomContrast} illustrates these oscillations of the contrast, damped over distances of the order of $c/\Gamma$.  More specifically, for large Rabi frequencies and non-negligible decay $\Gamma\tau_c$, the contrast approximates very well as
\begin{equation}
C=\e^{-\Gamma\tau_c/2}+\e^{-3\Gamma\tau_c/4}\cos(\Omega_M\tau_c).\label{eq:C1Om}
\end{equation}

%We note that for a negligible delay time ($\tau_c\ll \Omega_M^{-1},\Gamma^{-1}$), Eq.(\ref{eq:I1ate}) reduces to
%\begin{eqnarray}
%\frac{I(\mathbf{k},t)}{I_a}&=&\frac{s}{1+s}[1+\cos(2kz\cos\theta)],\label{no_tau}
%\end{eqnarray}
%in which case the saturation does not affect the single-atom fringe contrast. Indeed, each frequency scattered by the atom and its mirror image have the same phase at any time due to their proximity.

\section{\textsc{m}CBS in large clouds\label{sec:3}} 

The scenario may change dramatically in large disordered clouds, since the random phase acquired by the atoms from the laser may blur the fringes. The saturation of the atoms is expected to contribute further to the decrease of the contrast, since the Rabi frequency seen by an atom depends on its position in the standing wave, giving rise to different fluorescence spectra for atoms at different positions.

The electric field of the scattered light is now the superposition of the field scattered by all atoms, each one at a position $\mathbf{r}_j = (x_j,y_j,z_j)$, with $j \in \{1,...,N\}$ which indicates each one of the $N$ atoms. Let us discuss the case of optically dilute clouds, where the light-mediated dipole-dipole interaction can be neglected. The radiation of each atom is then described by single scattering theory, with an electric field exactly as in the single atom case \eqref{eq:Et}. The total light intensity has thus now the form
\begin{eqnarray}
&& \frac{I(\mathbf{k},t)}{I_a}=2\sum_{j,m}\langle\hat\sigma_j^\dagger(t)\hat\sigma_m(t)\rangle \e^{i\mathbf{k}_\perp\cdot(\mathbf{r}_j-\mathbf{r}_m)}\cos[k\cos\theta(z_j-z_m)]\nonumber
\\ && +\sum_{j,m}\langle\hat\sigma_j^\dagger(t)\hat\sigma_m(t+\tau_c)\rangle \e^{i(\mathbf{k}\cdot\mathbf{r}_j-\mathbf{k}'\cdot\mathbf{r}_m)}+\text{c.c.},\label{eq:I2gen}
\end{eqnarray}
where $\mathbf{k}_\perp=k (\sin \theta \cos \varphi, \sin \theta \sin \varphi,0)$.
We have made here the approximation that the dispersion between the different atoms of the cloud is negligible, i.e., $(\hat{\mathbf{n}}' \cdot\mathbf{r}_j -\hat{\mathbf{n}}\cdot\mathbf{r}_m)/c \cong \tau_c$ for all $j$ and $m$. This approximation is well justified for cold atoms experiments with atomic clouds at most centimeter-sized, and driving Rabi frequencies of hundreds of MHz.
%From Eq.\eqref{eq:I2gen} the appearance of two-time correlators can be noted. Over the size of a cloud, the time delay between the emission of different photons is negligible, nevertheless the travel time to the mirror, and back, introduces delays which may spread in phase the different spectral component of the light scattered by a strongly driven atom. In order to understand well the role of such decoherence in large clouds, let us first study the case of resonant light incident on a single atom in front of a mirror.
%For a single atom the sum \eqref{eq:I2gen} simplifies into
In the single scattering theory, two-atom connected correlations are null, which in the steady-state reads:
\begin{multline}
\langle\hat\sigma_j^\dagger(t)\hat\sigma_m(t')\rangle=\langle\hat\sigma_j^\dagger(t)\rangle\langle\hat\sigma_m(t')\rangle=\langle\hat\sigma_j^\dagger(t)\rangle\langle\hat\sigma_m(t)\rangle,\\
\forall t, t'\text{ and for } m\neq j,
\end{multline}
so the intensity decomposes as
\begin{multline}
\frac{I(\mathbf{k},t)}{I_a}= 2\sum_{j}\langle\hat\sigma_j^\dagger(t)\hat\sigma_j(t)\rangle\label{eq:I2N} \\
+2\sum_{j}\Re\left(\langle\hat\sigma_j^\dagger(t)\hat\sigma_j(t+\tau_c)\rangle \e^{-2ikz_j\cos\theta }\right) \\ 
+4\Re\left\{\sum_{j,m\neq j}\langle\hat\sigma_j^\dagger(t)\rangle\langle\hat\sigma_m(t)\rangle \e^{i\mathbf{k}_\perp\cdot(\mathbf{r}_j-\mathbf{r}_m)}\cos[k\cos\theta(z_j-z_m)]\right\}.
\end{multline}
The two first lines describe single-atom contributions, which can be obtained by summing \eqref{eq:I1ate} over the different atoms and their different saturation parameters, while the last term stands for the interference between them. %Again, we consider that the delay time $\tau_c$ is the same for all the atoms of the clouds; this approximation is valid for cloud sizes much smaller than $c/\Omega_0$ (which is of the order of dozens of meters for MhZ Rabi frequencies).

Averaging over the (uncorrelated) disorder of the cloud makes the last sum in Eq.\eqref{eq:I2N} disappear due to the vanishing average of the transverse phase term $\exp(i\mathbf{k}_\perp\cdot(\mathbf{r}_j-\mathbf{r}_m))$, so the intensity is the sum of the single-atom ones:
\begin{equation}
\frac{I(\mathbf{k},t)}{I_a}=\sum_j\frac{s_j}{1+s_j}+2\sum_j\Re\left(\langle\hat\sigma_j^\dagger(t)\hat\sigma_j(t+\tau_c)\rangle e^{-2ikz_j \cos \theta}\right).\label{eq:I2s}
\end{equation}

The first term in \eqref{eq:I2s} provides an isotropic background, with no dependence on $\mathbf{k}$, as one generally expects from an average over disorder. We note here that in the far-field limit the transverse dimensions ($x$ and $y$) of the cloud play no role when summing the contributions of the atoms. Switching to the continuous limit, we assume a Gaussian atomic density: 
\begin{equation}
\rho(\mathbf{r})=\frac{N}{(2\pi)^{3/2}\sigma_x\sigma_y\sigma_z}\exp\left(-\frac{x^2}{2\sigma_x^2}-\frac{y^2}{2\sigma_y^2}-\frac{(z+h)^2}{2\sigma_z^2}\right)\nonumber
\end{equation}
with $h$ the average distance of the cloud to the mirror. The saturation parameter at resonance for each atom is, as before, $s(z) = 2|\Omega (z)|^2/\Gamma^2 = s_0\cos^2(kz\cos\theta_0)$, and thus only the integral over $z$ gives a non-trivial result. We obtain
\begin{eqnarray}
\frac{I(\mathbf{k},\tau_c)}{I_a} &=&\frac{N}{\sqrt{2\pi}\sigma_z}\int dz\, \e^{-(z+h)^2/2\sigma_z^2}\Big[\frac{s(z)}{1+s(z)}\nonumber
\\ &&+2\cos(2kz\cos\theta)\left\langle\hat\sigma^\dagger\left(t\right)\hat\sigma\left(t+\tau_c\right)\right\rangle_z\Big].\label{eq:I2sb}
\end{eqnarray}
where the subscript $z$ in the two-times correlator indicates that we must consider the local Rabi frequency.

\subsection{Mirror close to the cloud}

In a previous work~\cite{Moriya2016} Eqs.\eqref{eq:I2s} and \eqref{eq:I2sb} had been derived, but without the two-time correlator which reflects the finite coherence time in the system: In the $\tau_c\to0$ limit, when the mirror is close to the atomic cloud, the results of that previous work are recovered. 

In particular, it was shown that in the low drive regime ($\Omega_0\ll\Gamma$) where the inelastic scattering contribution is negligible, the standing wave creates a grating of excited population of step $\lambda\cos\theta_0$ as the atoms respond linearly to the incident field. Let us remind that we assume there is no density modulation, only a Gaussian shape for the cloud.  The grating of excited population then produces a constructive interference in directions $\theta=\theta_0 \mod(\pi/kh\theta_0)$, in an angular opening of $1/2\theta_0k\sigma_z$ around $\theta_0$ (see Eqs.~(\ref{eq:linear}--\ref{eq:ftheta})).  One can then show that the intensity reads
\begin{equation}
\frac{I(\theta)}{I_a}=N \frac{s_0}{2}\left[1+\frac{1}{2}f(\theta)\right] \label{eq:linear}
\end{equation}
with
\begin{equation}
f(\theta)=\e^{-2(\theta_0 k\sigma_z)^2(\theta-\theta_0)^2}\cos(2\theta_0 kh(\theta-\theta_0)).\label{eq:ftheta}
\end{equation}
We see that, as in the case of CBS, where the cloud density can affect the shape of the CBS cone~\cite{Labeyrie2003}, the envelope of the mCBS interference fringes also depends on the cloud spatial density.
%Note that the cloud density may affect the shape of the CBS cone, in the case of the mCBS it will not affect the fringes, but rather their envelop. 
Now, as we increase the saturation parameter for a cloud close to the mirror, the mCBS fringes obtained from the modulation of excited population decreases monotonically since the atomic population saturates everywhere apart from a vanishing region around the nodes of the standing wave $2\Omega_0\cos(kz\cos\theta_0)$, as deduced in Ref.~\cite{Moriya2016}. 
%For increasing pump intensity, this region gets smaller as the local Rabi frequency puts most atoms in the saturated regime. 

\subsection{Mirror far from the cloud}

\begin{figure}[!h]
%\begin{center}
\includegraphics[scale=0.65]{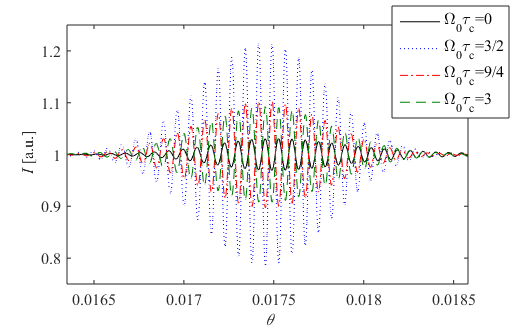}
%\end{center}
%\vspace{-2.05cm}
\caption{\label{fig:vartaud} Intensity pattern for a large cloud in the highly saturated regime for different time delay times $\tau_c$. Simulations realized for a cloud with $\sigma_z=1$cm and $\Omega_0=10\Gamma$, with a laser incidence angle of $1^\circ$.}
\end{figure}

There is nevertheless another mechanism that can maintain constructive interference in the saturated regime, namely the role of the sidebands, as revealed by a closer analysis of the two-time correlator~\eqref{eq:2tc}. To pin down this effect, let us first neglect the inhomogeneity in the excited population of the atomic cloud, as it is particularly relevant for high saturation parameters. Practically, we assume that in Eqs.\eqref{eq:2tc} and \eqref{eq:I2sb}, the elastic scattering contribution can be neglected ($s/2(1+s)^2\approx0$), as well as the modulation of the excited population ($s/(1+s)\approx1$); we also assume $(s-1)/(s+1)\approx 1$ to neglect the extra modulation of the Mollow sidebands, and the last term in \eqref{eq:2tc}, which scales as $\Gamma/\Omega_M\sim 1/\sqrt{s}$, is also neglected. The expression of the intensity \eqref{eq:I2sb} then reduces to
\begin{eqnarray}
&&\frac{I(\mathbf{k},\tau_c)}{I_a}=\frac{N}{\sqrt{2\pi}\sigma_z}\int dz\, \e^{-(z+h)^2/2\sigma_z^2}\Big[1+ \label{eq:IMollow}
\\ &&\,\,\frac{\cos(2kz\cos\theta)}{2}\left(\e^{-\Gamma \tau_c/2}+\cos(\Omega_M (z)\tau_c)\e^{-3\Gamma\tau_c/4}\right)\Big].\nonumber
\end{eqnarray}
Instead of the excited population, the gradient in the atomic cloud that gives rise to the interference pattern now originates in the modulated Rabi frequency $\Omega_M (z)$ in the last term of Eq. \eqref{eq:IMollow}, whereas the resonant peak term, proportional to $\e^{-\Gamma \tau_c/2}$, will clearly have a vanishing average for large clouds. 
%Due to the delay $\tau_c$, the interference of the two Mollow sidebands, positively and negatively shifted with respect to the atomic transition by $\Omega_M(z)$, results in a periodic modulation in space of the intensity emitted by these sidebands, and thus in coherent emission for specific angles. 
The resulting fringes depend on the delay time $\tau_c$, as it can be seen in Fig.~\ref{fig:vartaud}.

\begin{figure}[!h]
\begin{center}
\includegraphics[scale=0.52]{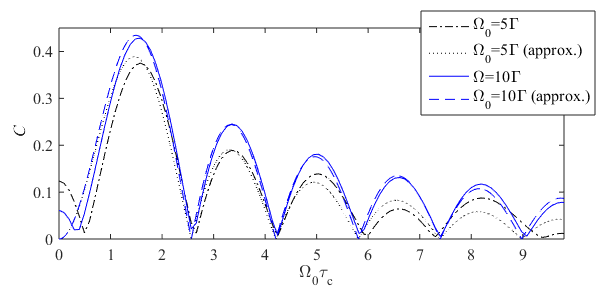}
\end{center}
%\vspace{-2.05cm}
\caption{\label{fig:contrast}Constrast of the fringes as a function of the product $\Omega_0\tau_c$. Simulations realized for a Gaussian cloud of $\sigma_z=1$cm at $30$cm from the mirror, illuminated by a plane wave with Rabi frequency $\Omega_0=5\Gamma$ and $\Omega_0=10\Gamma$ and inclination angle $\theta_0=1^\circ$. The dash-dotted and plain curve refer to full expressions \eqref{eq:2tc} and \eqref{eq:I2s}, whereas the dotted and dashed ones to the approximate expression \eqref{eq:Capp}.}
\end{figure}

The contribution of the spatial modulation of the Rabi frequency can be more precisely evaluated by approximating the local Rabi frequency as $\Omega_M(z)\approx \Omega(z) =2\Omega_0\cos(kz\cos\theta_0)$ and expanding it in Fourier modes
\begin{eqnarray}
\cos[2\Omega_0\tau_c\cos(kz&&\cos\theta_0)]=J_0(2\Omega_0\tau_c)+
\\ &&2\sum_{n=1}^\infty (-1)^n J_{2n}(2\Omega_0\tau_c)\cos(2nkz\cos\theta_0).\nonumber
\end{eqnarray}
where the $J_{\alpha}(z)$ are the Bessel functions of first kind. We thus see that inside the integral \eqref{eq:IMollow} there will be an infinite series of terms of the form $\cos(2kz\cos\theta)\cos(2nkz\cos\theta_0)$. For large clouds, all these terms average out to zero, except for the $n=1$ term at observation angles $\theta\approx\theta_0\ll 1$. Then the intensity simplifies into
\begin{equation}
\frac{I(\mathbf{k},\tau_c)}{N\, I_a}=1-\frac{J_2(2\Omega_0\tau_c)}{2}e^{-3\Gamma\tau_c/4}f(\theta).\label{eq:IMollowf}
\end{equation}
This results in a contrast:
\begin{equation}
C\approx |J_2(2\Omega_0\tau_c)| e^{-3\Gamma\tau_c/4}.\label{eq:Capp}
\end{equation}
At large Rabi frequencies, the above formula is in very good agreement with the contrast extracted from the exact value of the two-time correlator \eqref{eq:2tc}, as can be observed in Fig.~\ref{fig:contrast}. Interestingly, the maximum contrast is reached for $\Omega_0\tau_c\approx 3/2$, rather than at $\tau_c=0$ where the absence of delay does not allow for the Mollow sidebands to produce a modulation of the emission (see Fig. \ref{fig:contrast}). Thus introducing a substantial distance between the mirror and the atoms allows to observe fringes which originate in the Mollow sidebands: The mCBS setup allows to observe interferences based on inelastically scattered photons only, with a contrast which depends directly on the Mollow spectrum. Moreover, by tuning the position of the mirror and thus the parameter $\tau_c$, the dependence of the contrast on the Rabi frequency can be varied in such a way that one can obtain a contrast larger than $0.4$ for an arbitrarily high saturation of the atomic transition.

\section{mCBS spectrum\label{sec:4}} 

\begin{figure}[!h]
\begin{center}
\includegraphics[scale=0.7]{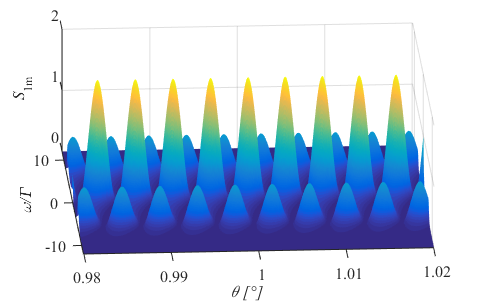}
\end{center}
%\vspace{-2.05cm}
\caption{Emitted spectrum of a single atom in front of a mirror $S_{1\text{m}}(\omega)$ (see Eq.~\eqref{eq:S1m}), as a function of observation angle $\theta$. Simulations realized for an atom at $0.3$m from the mirror, illuminated by a plane-wave with Rabi frequency $\Omega_0=3\Gamma$ and incidence angle $\theta_0=1^\circ$, and $\Omega_0\tau_c=\pi/2$.}\label{fig:radiation1atom}
\end{figure}

Let us have a closer look at the spectral features of the reflected light, to confirm the specific role of the Mollow sidebands. The radiation spectrum of the cloud is given by the Fourier transform of the first-order optical coherence
\begin{equation}
S(\omega)=\int_{-\infty}^{\infty} d\tau e^{-i\omega\tau}\lim_{t\to\infty}\langle\hat E^\dagger(t)\hat E(t+\tau)\rangle.
\end{equation}
%\begin{equation}
%g^{(1)}(\tau)=\lim_{t\rightarrow\infty}\frac{\langle\hat E^\dagger(t)\hat E(t+\tau)\rangle}{\langle\hat E^\dagger\hat E\rangle}
%\end{equation}
%by computing its Fourier transform:
%\begin{eqnarray}
%S(\omega)&=&\int_{-\infty}^{\infty}g^{(1)}(\tau) e^{-i\omega\tau}d\tau
%\\ &=& 2 \mathrm{Re}\left(\int_{0}^{\infty}g^{(1)}(\tau) e^{-i\omega\tau}d\tau\right).\nonumber
%\end{eqnarray}
Note that since the different atoms are not driven by the same Rabi frequency, we did not normalize the optical coherence by the usual term $\langle\hat E^\dagger(t)\hat E(t)\rangle$~\cite{Scully1997}. Returning to the case of a single atom in front of the mirror, the mCBS fluorescence spectrum $S_{1\text{m}}$ of a single atom at position $\mathbf{r}$ is derived from Eqs.\eqref{eq:Et} and \eqref{eq:2tc}:
\begin{equation}
S_{1\text{m}}(\mathbf{r},\omega,\theta)=2S_1(\mathbf{r},\omega)\left[1+\cos\left(2kz\cos\theta-\omega\tau_c\right)\right],\label{eq:S1m}
\end{equation}
where $S_1(\mathbf{r},\omega)$ refers to the single-atom spectrum in absence of a mirror and driven by a plane-wave with Rabi frequency $\Omega(z)$:
\begin{eqnarray}
&S_1(z,\omega)=\frac{\pi s(z)}{[1+s(z)]^2}\delta(\omega)+\frac{s(z)}{4[1+s(z)]}\mathrm{Re}\Bigg\{ \frac{1}{\Gamma/2+i\omega}\label{eq:S1nom}  \\
&+\frac{1}{2}\frac{s(z)-1}{s(z)+1}\left[\frac{1}{3\Gamma/4+i(\omega-\Omega_M(z))}+\frac{1}{3\Gamma/4+i(\omega+\Omega_M(z))}\right] \nonumber\\
&+\frac{\Gamma}{8i\Omega_M}\frac{5s(z)-1}{s(z)+1}\left[\frac{1}{3\Gamma/4+i(\omega-\Omega_M(z))}-\frac{1}{3\Gamma/4+i(\omega+\Omega_M(z))}\right]\Bigg\}.\nonumber
\end{eqnarray}
The mCBS physics lies in Eq.\eqref{eq:S1m}, where one observes that the mirror induces two sources of modulation of the single-atom spectrum: The created standing wave,  which modulates the Rabi frequency $\Omega(z)$, and the cosine interference term: The argument of the cosine, $2kz \cos \theta - \omega \tau_c$, makes that different frequencies $\omega$ will present different angular fringes' maxima $\theta$.
%; The presence of the $\omega\tau_c$ phase term makes that different frequencies will present different fringe positions. 
As a consequence, the emission spectrum of a single atom is a function of the emission angle $\theta$ and the atom position $z$, as can be clearly seen in Fig. \ref{fig:radiation1atom}. Depending on its position in the standing wave and on the delay term $\tau_c$, the interference of the light directly scattered at the angle $\theta$ with the light reflected at $\theta$ by the mirror can 
cancel the contribution of certain frequencies and amplify others, as shown by the cosine in Eq.\eqref{eq:S1m}. For example, atoms at a maximum of the standing wave ($kz\cos\theta_0=0\mod\pi$) will always have a maximum resonant ($\omega\approx 0$) emission in the $\theta_0$ angle, yet the emission of the Mollow sidebands in this direction are canceled for $\Omega_M\tau_c=\pi\mod(2\pi)$. On the other hand, for same $\Omega_M \tau_c$, the same atom will present an opposite scenario in a different angle, where the resonant emission cancels and the Mollow sidebands are amplified by the presence of the mirror (specifically, for $\theta$ such that $kz\cos\theta=\pi\mod 2\pi$). Fig.\ref{fig:radiation1atom} illustrates this effect, where the interference of the radiation from the atom and its mirror image is observed to be constructive at different angles for the resonant frequency and the Mollow sidebands.
%For an optically dilute large cloud, the single scattering hypothesis means that the total spectrum will be given by the sum of all the single atom spectra over the atomic cloud. Addressing the dipole-dipole interaction in optically thick clouds requires treating connected correlations, which may bring new fluorescence peaks [Mollow collectif, etc]; it is however beyond the scope of the present paper.

\begin{figure}[!h]
\begin{center}
\includegraphics[scale=0.75]{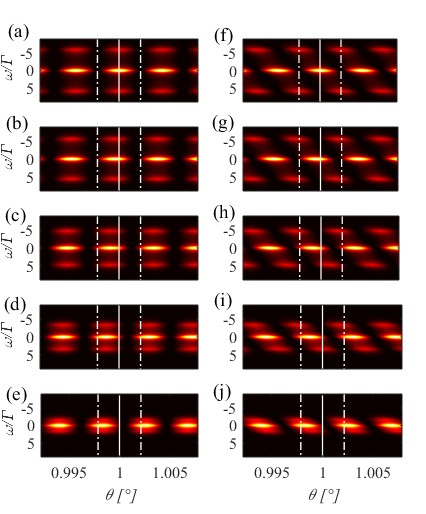}
\end{center}
%\vspace{-2.05cm}
\caption{Emitted spectrum of a single atom in front of a mirror, $S_{1\text{m}}(\omega)$ of eq.~(\ref{eq:S1m}), as a function of observation angle $\theta$, for different positions $\mathbf{r}$ in the standing wave. Panels (a-e) correspond to the condition $\tau_c=0$, and panels (f-j) to $\tau_c\sqrt{4\Omega_0^2-\Gamma^2/16}=\pi$. The five atoms are located at $z=-h+\delta z$, with $k\delta z=0,\ \pi/10,\ 2\pi/10,\ 3\pi/10,\ 4\pi/10$, from top to bottom. Simulation realized with a plane-wave of incidence angle $1^\circ$ and Rabi frequency $3\Gamma$, for a cloud of length $1$mm and at $30$cm from the mirror.}\label{fig:radiation1atomd}
\end{figure}

Returning to large disordered clouds, we use the same hypotheses for the saturated regime as before: The elastically scattered term is dropped ($s(z)/2[1+s(z)]^2\ll 1$), and the spatial intensity modulation of the standing wave created by the incident and reflected laser beams is neglected ($s(z)/(1+s(z))\approx (s(z)-1)/(s(z)+1)\approx 1$).
The cloud spectrum is then obtained by summing the contributions of the three inelastic peaks $S(\omega)=S_0(\omega)+S_+(\omega)+S_-(\omega)$, where
\begin{eqnarray}
S_0(\omega) &=& \frac{\Gamma}{8} \int d\mathbf{r} \rho(z)\frac{1+\cos(2kz\cos\theta-\omega\tau_c)}{\omega^2+(\Gamma/2)^2}\label{eq:S0}
\\ S_\pm(\omega) &=& \frac{3\Gamma}{32} \int d\mathbf{r} \rho(z)\frac{1+\cos(2kz\cos\theta-\omega\tau_c)}{[\omega\mp\Omega(z)]^2+(3\Gamma/4)^2}\label{eq:Spm}
\end{eqnarray}
and $\Omega(z)=2\Omega_0\cos(kz\cos\theta_0)$.
The absence of spatial modulation in $S_0$ for the radiation of the resonant light ($\omega\approx 0$) makes that, due to the cosine term in eq.~(\ref{eq:S0}), atoms with different $z$ position coordinates will have fringe patterns which are shifted in $\theta$, so averaging over the cloud results in the disappearance of these fringes and in a mere background radiation. In the case $\tau_c=0$, the Mollow sidebands present the same behaviour. This effect is illustrated in Fig.~\ref{fig:radiation1atomd}(a-e), where the angular dependence of the spectra of atoms at different positions of the standing wave exhibit maxima of the resonant emission at different angles, and the sum over all positions of a large cloud (much larger than $2\pi/k$) shows no fringe. %Consequently we have $S_0(\omega)=(\Gamma/8)/(\omega^2+\Gamma^2/4)$.

On the contrary, for finite delay times $\tau_c$, the situation can be quite different. We show in Fig.\ref{fig:radiation1atomd}(f-j) the angular dependence of the atomic spectra for different atomic positions, for $\tau_c \sqrt{4 \Omega_0^2 - \Gamma^2/16} = \pi$. In this case, the upper Mollow sideband appear to produce fringes in an almost-constant angle, independent of the atomic position at the cloud.
%On the contrary, when the delay time $\tau_c$ is appropriately chosen, despite the resonant peak still present a strong angular dependence (and thus no fringes after disorder averaging), the Mollow sidebands appear to produce fringes in an almost-constant angle, see Fig.~\ref{fig:radiation1atomd}(f-j). 
Consequently, the total intensity resulting from the average over the many atoms of the cloud will still present fringes, above the background of the resonant peak.

%The emission from the Mollow sidebands, on the contrary, is spatially modulated, which may result in directional emission and fringes. %Specifically, the central frequency of each sideband is a function of the atomic position within the laser standing wave through the Rabi frequency dependence on the laser intensity, $\Omega(\mathbf{r})$, as it is evident from Eq.~(\ref{eq:S1nom}). Because of that, the calculation of the angular maxima of the fringes created by each one of the sidebands of the spectrum of a specific atom is different than for the resonant emission.
%, as it can be seen from the phase of the cosine term of Eq.~(\ref{eq:S1m}): Now, the emission directions $\theta$ that maximize the cosine term for each atom will be a function of their position $z$ not only through the phase term $2kz\cos\theta$, but also through the $z$ dependence of their central frequency $\omega = \pm \Omega_M(\mathbf{r}) = \pm \Omega_M(z)$ of each sideband. 
%For specific values of the Rabi frequency $\Omega_0$ of a single laser beam, both spatial modulations of the Rabi frequency and of the angular emission profile can become approximately commensurate, in such a way that, after disorder-averaging, an angular dependence of the fringes still remains. 
For a yet simpler picture to characterize the directional emission of the Mollow triplet 
%In order to characterize the directional emission from the Mollow triplet 
we use the following property to describe the two sidebands in Eq.\eqref{eq:Spm} 
\begin{equation}
\lim_{\gamma\to 0}\frac{\gamma}{x^2+\gamma^2}=\pi\delta(x),\label{eq:dirac}
\end{equation}
with $\delta$ is the Dirac delta function, $\gamma=3\Gamma/4\Omega_0$ and $x=\omega/\Omega_0$. We are thus describing the two sidebands as Dirac functions in frequency for the highly saturated regime, assuming $\Omega_0/\Gamma\gg 1$. A continuous spectrum is still obtained due to the range of Rabi frequency $\Omega(z)$ present in the system. After some calculations presented in Appendix~\ref{appx:spec}, the inelastic spectrum can be computed as
\begin{eqnarray}
S(\omega)&=&\frac{\Gamma/8}{\omega^2+\Gamma^2/4}+\frac{1}{4\sqrt{4\Omega_0^2-\omega^2}}\label{eq:SpD}
\\ & & \times \left[1+\e^{-2(\theta_0 k\sigma_z)^2(\theta-\theta_0)^2}\cos(2\theta_0 kh(\theta-\theta_0)-\omega\tau_c)\right],\nonumber
\end{eqnarray}
which is defined for $|\omega|<2\Omega_0$ due to the Dirac assumption for each sideband. Eq.\eqref{eq:SpD} shows clearly that the fringes originate in the Mollow sidebands. Furthermore, the central inelastic peak \eqref{eq:S0} presents the usual Lorentzian shape, whereas the sidebands exhibit a $1/\omega$ decay that reflects the spread in Mollow frequencies due to the inhomogeneous intensity profile. One remarks that for any $\tau_c$ a specific spectral component of the sidebands presents fringes (the central peak does not) yet integrating over the whole spectrum yields
\begin{equation}
I(\theta)\sim 1+\frac{J_0(2\Omega_0\tau_c)}{2}f(\theta).\label{eq:ContrastD}
\end{equation}
Thus only specific delay times $\tau_c$ yield an optimal contrast for the total intensity, i.e., the values of $\Omega_0\tau_c$ which correspond to maxima of the $J_0$ function. Let us remark that $J_0(w)\approx -J_2(w)$ for $w\gg 1$, which makes Eq.\eqref{eq:ContrastD} compatible with Eq.\eqref{eq:C1Om}, up to the $\e^{-3\Gamma\tau_c/4}$ term which reflects the finite linewidth of the sidebands in Eq.\eqref{strong_s2}.

Fig.~\ref{fig:radiation1atomd} shows this effect for a value of $\Omega_0$ such that $\tau_c\sqrt{4\Omega_0^2-\Gamma^2/16}=\pi$, which corresponds to high contrast fringes for the saturated regime. In this condition, one observes that the maxima of the fringes created by the upper Mollow sideband are always around the angles $\theta_0+\pi/2\theta_0kh\mod(\pi\theta_0kh)$. Because of the commensurability of the two modulations, the lower Mollow sideband will also give after disorder-averaging a non-zero contribution to the contrast, since the spatial modulation of $2kz\cos\theta$ is linear, while that of $\Omega_M(z)$ is (approximately) sinusoidal. This spectral analysis confirms that the fringes observed over the background are composed of light scattered inelastically into the two Mollow sidebands.

\section{Conclusion\label{sec:5}}

We have here discussed the emergence of coherent backscattering of light by large clouds of saturated atoms in presence of a mirror. We have shown that, at odds with interference effects based on multiple scattering, or on interference between different scatterers~\cite{Wolf2016}, the presence of a mirror allows for the fringes' contrast to survive even in the strong saturation limit. Moreover, disorder-averaging of the fluorescence does not destroy the fringes, which makes mCBS a robust, scalable platform for probing temporal correlations of the light radiated  by strongly driven correlated scatterers. This system can be used as a valuable tool for detecting deviations from the Mollow theory, such as those caused by modifications on the electromagnetic vacuum surrounding the scatterers~\cite{Toyli2016}, by an enhancement or suppression of vacuum modes (as when the scatterers are coupled to a cavity~\cite{Kim2014}), or by collective effects on the saturated scattering of light from the atomic cloud~\cite{Pucci2017}.

%This spatial mapping of the spectral properties of the light scattered by disordered atoms represents a simple and robust system for probing temporal correlations and spectral properties of the light scattered by strongly-driven correlated scatterers.

\appendix

\begin{widetext}

\section{Derivation of Eq.\eqref{eq:SpD}}
\label{appx:spec}

Considering the limit $\Omega_0\gg\Gamma$, one can use \eqref{eq:dirac} onto \eqref{eq:Spm} to obtain
\begin{eqnarray}
S_\pm(x)&=&\frac{\pi}{8\Omega_0}\int dz\rho(z)\left[1+\cos(2kz\cos\theta-\alpha x)\right]
\delta[x\mp 2\cos(kz\cos\theta_0)]
\end{eqnarray}
where $\alpha=\Omega_0\tau_c$ and $x=\omega/\Omega_0$.
The argument of the Dirac delta function in the above equation can be seen as a function of $z$  for a given value of the dimensionless frequency $x$, i.e. 
\begin{equation}
\delta[x\mp 2\cos(kz\cos\theta_0)]=\delta[f_{\pm}(z)]=\sum_j\frac{\delta(z-z_j)}{|f_\pm'(z_j)|}
\end{equation}
where $f_\pm(z)=2\cos(kz\cos\theta_0)\mp x$ and
$z_j$ are its zeros, i.e., the solutions of the equation $\cos(kz_j\cos\theta_0)=\pm x/2$ for $|x|\le 2$, and $|f'(z_j)|=2k|\cos\theta_0|\sqrt{1-x^2/4}$. The $z_j$ for the upper and lower sideband are $kz_j\cos\theta_0=\pm\arccos(x/2)+2\pi j$ and $kz_j\cos\theta_0=\pm\arccos(x/2)+\pi(2j+1)$ respectively, with $j\in \mathbb{Z}$. 
%for an infinite density such as the Gaussian under consideration. 
By integrating the Dirac function leads to:
\begin{eqnarray}
S_+(x)&=& \frac{\pi}{16k\Omega_0|\cos\theta_0|\sqrt{1-x^2/4}}\sum_j\left\{\left[1+\cos\left((2\arccos(x/2)+4\pi j)\frac{\cos\theta}{\cos\theta_0}-\alpha x\right)\right]\rho\left(\frac{\arccos(x/2)+2\pi j}{k\cos\theta_0}\right)\right.\nonumber\\
&+& \left.\left[1+\cos\left((2\arccos(x/2)+4\pi j)\frac{\cos\theta}{\cos\theta_0}+\alpha x\right)\right]\rho\left(\frac{-\arccos(x/2)-2\pi j}{k\cos\theta_0}\right)\right\}\\
S_-(x)&=& \frac{\pi}{16k\Omega_0|\cos\theta_0|\sqrt{1-x^2/4}}\sum_j\left\{\left[1+\cos\left((2\arccos(x/2)+2\pi(2j+1))\frac{\cos\theta}{\cos\theta_0}-\alpha x\right)\right]\rho\left(\frac{\arccos(x/2)+\pi(2j+1)}{k\cos\theta_0}\right)\right.\nonumber\\
&+& \left.\left[1+\cos\left((2\arccos(x/2)+2\pi(2j+1))\frac{\cos\theta}{\cos\theta_0}+\alpha x\right)\right]\rho\left(\frac{-\arccos(x/2)-\pi(2j+1)}{k\cos\theta_0}\right)\right\}.
\end{eqnarray}
Considering small angles $\theta,\theta_0\ll 1$, we use that $\cos\theta/\cos\theta_0\approx 1-\theta_0(\theta-\theta_0)$ to write that
\begin{eqnarray}
\cos\left((2\arccos(x/2)+2\pi m)\frac{\cos\theta}{\cos\theta_0}\mp\alpha x\right)
&=&\cos(2\arccos(x/2)+2\pi m\mp\alpha x)\cos[(2\arccos(x/2)+2\pi m)\theta_0(\theta-\theta_0))]\nonumber\\
&+&\sin(2\arccos(x/2)+2\pi m\mp\alpha x)\sin[(2\arccos(x/2)+2\pi m)\theta_0(\theta-\theta_0))]\nonumber
\end{eqnarray} 
Since $\theta_0(\theta-\theta_0)$ is a very small quantity and $\arccos(x/2)\le\pi$, fringes are observable only if $m\gg 1$, so the above equation simplifies into
\begin{equation}
\cos\left((2\arccos(x/2)+2\pi m)\frac{\cos\theta}{\cos\theta_0}\mp\alpha x\right)
=\cos(\alpha x\pm 2\pi m\theta_0(\theta-\theta_0)],
\end{equation}
and the Mollow sidebands contribution is equal to
\begin{equation}
S_+(x)+S_-(x)=\frac{\pi}{4k\Omega_0\sqrt{1-x^2/4}}\sum_j\left\{1+\cos[\alpha x+4\pi j\theta_0(\theta-\theta_0)]\right\}\rho\left(\frac{2\pi j}{k}\right).
\end{equation}
For a Gaussian distribution, the density term writes $\rho(2\pi j/k)=\exp(-(2\pi j+kh)^2/2(k\sigma_z)^2)/\sqrt{2\pi}\sigma_z$, and for large clouds ($k\sigma_z\gg 1$) the sum can be turned into an integral which yields
\begin{equation}
S_+(x)+S_-(x)=\frac{1}{8k\Omega_0\sqrt{1-x^2/4}}
\left[1+
e^{-2(k\sigma_z\theta_0)^2(\theta-\theta_0)^2}\cos(2kh\theta_0(\theta-\theta_0)-\alpha x)\right],
\end{equation}
which leads to Eq.\eqref{eq:SpD}.
\end{widetext}

\bibliography{../../../Biblio/BiblioCollectiveScattering}

\begin{thebibliography}{41}%
\makeatletter
\providecommand \@ifxundefined [1]{%
 \@ifx{#1\undefined}
}%
\providecommand \@ifnum [1]{%
 \ifnum #1\expandafter \@firstoftwo
 \else \expandafter \@secondoftwo
 \fi
}%
\providecommand \@ifx [1]{%
 \ifx #1\expandafter \@firstoftwo
 \else \expandafter \@secondoftwo
 \fi
}%
\providecommand \natexlab [1]{#1}%
\providecommand \enquote  [1]{``#1''}%
\providecommand \bibnamefont  [1]{#1}%
\providecommand \bibfnamefont [1]{#1}%
\providecommand \citenamefont [1]{#1}%
\providecommand \href@noop [0]{\@secondoftwo}%
\providecommand \href [0]{\begingroup \@sanitize@url \@href}%
\providecommand \@href[1]{\@@startlink{#1}\@@href}%
\providecommand \@@href[1]{\endgroup#1\@@endlink}%
\providecommand \@sanitize@url [0]{\catcode `\\12\catcode `\$12\catcode
  `\&12\catcode `\#12\catcode `\^12\catcode `\_12\catcode `\%12\relax}%
\providecommand \@@startlink[1]{}%
\providecommand \@@endlink[0]{}%
\providecommand \url  [0]{\begingroup\@sanitize@url \@url }%
\providecommand \@url [1]{\endgroup\@href {#1}{\urlprefix }}%
\providecommand \urlprefix  [0]{URL }%
\providecommand \Eprint [0]{\href }%
\providecommand \doibase [0]{http://dx.doi.org/}%
\providecommand \selectlanguage [0]{\@gobble}%
\providecommand \bibinfo  [0]{\@secondoftwo}%
\providecommand \bibfield  [0]{\@secondoftwo}%
\providecommand \translation [1]{[#1]}%
\providecommand \BibitemOpen [0]{}%
\providecommand \bibitemStop [0]{}%
\providecommand \bibitemNoStop [0]{.\EOS\space}%
\providecommand \EOS [0]{\spacefactor3000\relax}%
\providecommand \BibitemShut  [1]{\csname bibitem#1\endcsname}%
\let\auto@bib@innerbib\@empty
%</preamble>
\bibitem [{\citenamefont {Wolf}\ \emph {et~al.}(2016)\citenamefont {Wolf},
  \citenamefont {Wechs}, \citenamefont {von Zanthier},\ and\ \citenamefont
  {Schmidt-Kaler}}]{Wolf2016}%
  \BibitemOpen
  \bibfield  {author} {\bibinfo {author} {\bibfnamefont {S.}~\bibnamefont
  {Wolf}}, \bibinfo {author} {\bibfnamefont {J.}~\bibnamefont {Wechs}},
  \bibinfo {author} {\bibfnamefont {J.}~\bibnamefont {von Zanthier}}, \ and\
  \bibinfo {author} {\bibfnamefont {F.}~\bibnamefont {Schmidt-Kaler}},\
  }\bibfield  {title} {\enquote {\bibinfo {title} {Visibility of young's
  interference fringes: Scattered light from small ion crystals},}\ }\href@noop
  {} {\bibfield  {journal} {\bibinfo  {journal} {Phys. Rev. Lett.}\ }\textbf
  {\bibinfo {volume} {116}},\ \bibinfo {pages} {183002} (\bibinfo {year}
  {2016})}\BibitemShut {NoStop}%
\bibitem [{\citenamefont {Kuga}\ and\ \citenamefont
  {Ishimaru}(1984)}]{Kuga1984}%
  \BibitemOpen
  \bibfield  {author} {\bibinfo {author} {\bibfnamefont {Yasuo}\ \bibnamefont
  {Kuga}}\ and\ \bibinfo {author} {\bibfnamefont {Akira}\ \bibnamefont
  {Ishimaru}},\ }\bibfield  {title} {\enquote {\bibinfo {title}
  {Retroreflectance from a dense distribution of spherical particles},}\ }\href
  {\doibase 10.1364/josaa.1.000831} {\bibfield  {journal} {\bibinfo  {journal}
  {Journal of the Optical Society of America A}\ }\textbf {\bibinfo {volume}
  {1}},\ \bibinfo {pages} {831} (\bibinfo {year} {1984})}\BibitemShut {NoStop}%
\bibitem [{\citenamefont {Albada}\ and\ \citenamefont
  {Lagendijk}(1985)}]{Albada1985}%
  \BibitemOpen
  \bibfield  {author} {\bibinfo {author} {\bibfnamefont {Meint P.~Van}\
  \bibnamefont {Albada}}\ and\ \bibinfo {author} {\bibfnamefont
  {Ad}~\bibnamefont {Lagendijk}},\ }\bibfield  {title} {\enquote {\bibinfo
  {title} {Observation of weak localization of light in a random medium},}\
  }\href {\doibase 10.1103/PhysRevLett.55.2692} {\bibfield  {journal} {\bibinfo
   {journal} {Phys. Rev. Lett.}\ }\textbf {\bibinfo {volume} {55}},\ \bibinfo
  {pages} {2692--2695} (\bibinfo {year} {1985})}\BibitemShut {NoStop}%
\bibitem [{\citenamefont {Wolf}\ and\ \citenamefont {Maret}(1985)}]{Wolf1985}%
  \BibitemOpen
  \bibfield  {author} {\bibinfo {author} {\bibfnamefont {Pierre-Etienne}\
  \bibnamefont {Wolf}}\ and\ \bibinfo {author} {\bibfnamefont {Georg}\
  \bibnamefont {Maret}},\ }\bibfield  {title} {\enquote {\bibinfo {title} {Weak
  localization and coherent backscattering of photons in disordered media},}\
  }\href {\doibase 10.1103/PhysRevLett.55.2696} {\bibfield  {journal} {\bibinfo
   {journal} {Phys. Rev. Lett.}\ }\textbf {\bibinfo {volume} {55}},\ \bibinfo
  {pages} {2696--2699} (\bibinfo {year} {1985})}\BibitemShut {NoStop}%
\bibitem [{\citenamefont {Yoo}\ \emph {et~al.}(1990)\citenamefont {Yoo},
  \citenamefont {Tang},\ and\ \citenamefont {Alfano}}]{Yoo1990}%
  \BibitemOpen
  \bibfield  {author} {\bibinfo {author} {\bibfnamefont {K.~M.}\ \bibnamefont
  {Yoo}}, \bibinfo {author} {\bibfnamefont {G.~C.}\ \bibnamefont {Tang}}, \
  and\ \bibinfo {author} {\bibfnamefont {R.~R.}\ \bibnamefont {Alfano}},\
  }\bibfield  {title} {\enquote {\bibinfo {title} {Coherent backscattering of
  light from biological tissues},}\ }\href {\doibase 10.1364/ao.29.003237}
  {\bibfield  {journal} {\bibinfo  {journal} {Applied Optics}\ }\textbf
  {\bibinfo {volume} {29}},\ \bibinfo {pages} {3237} (\bibinfo {year}
  {1990})}\BibitemShut {NoStop}%
\bibitem [{\citenamefont {Mishchenko}(1993)}]{Mishchenko1993}%
  \BibitemOpen
  \bibfield  {author} {\bibinfo {author} {\bibfnamefont {Michael~I.}\
  \bibnamefont {Mishchenko}},\ }\bibfield  {title} {\enquote {\bibinfo {title}
  {On the nature of the polarization opposition effect exhibited by saturns
  rings},}\ }\href {\doibase 10.1086/172835} {\bibfield  {journal} {\bibinfo
  {journal} {The Astrophysical Journal}\ }\textbf {\bibinfo {volume} {411}},\
  \bibinfo {pages} {351} (\bibinfo {year} {1993})}\BibitemShut {NoStop}%
\bibitem [{\citenamefont {Bayer}\ and\ \citenamefont
  {Niederdr\"{a}nk}(1993)}]{Bayer1993}%
  \BibitemOpen
  \bibfield  {author} {\bibinfo {author} {\bibfnamefont {G.}~\bibnamefont
  {Bayer}}\ and\ \bibinfo {author} {\bibfnamefont {T.}~\bibnamefont
  {Niederdr\"{a}nk}},\ }\bibfield  {title} {\enquote {\bibinfo {title} {Weak
  localization of acoustic waves in strongly scattering media},}\ }\href
  {\doibase 10.1103/physrevlett.70.3884} {\bibfield  {journal} {\bibinfo
  {journal} {Physical Review Letters}\ }\textbf {\bibinfo {volume} {70}},\
  \bibinfo {pages} {3884--3887} (\bibinfo {year} {1993})}\BibitemShut {NoStop}%
\bibitem [{\citenamefont {Wiersma}\ \emph {et~al.}(1995)\citenamefont
  {Wiersma}, \citenamefont {van Albada}, \citenamefont {van Tiggelen},\ and\
  \citenamefont {Lagendijk}}]{Wiersma1995}%
  \BibitemOpen
  \bibfield  {author} {\bibinfo {author} {\bibfnamefont {Diederik~S.}\
  \bibnamefont {Wiersma}}, \bibinfo {author} {\bibfnamefont {Meint~P.}\
  \bibnamefont {van Albada}}, \bibinfo {author} {\bibfnamefont {Bart~A.}\
  \bibnamefont {van Tiggelen}}, \ and\ \bibinfo {author} {\bibfnamefont
  {Ad}~\bibnamefont {Lagendijk}},\ }\bibfield  {title} {\enquote {\bibinfo
  {title} {Experimental evidence for recurrent multiple scattering events of
  light in disordered media},}\ }\href {\doibase 10.1103/physrevlett.74.4193}
  {\bibfield  {journal} {\bibinfo  {journal} {Physical Review Letters}\
  }\textbf {\bibinfo {volume} {74}},\ \bibinfo {pages} {4193--4196} (\bibinfo
  {year} {1995})}\BibitemShut {NoStop}%
\bibitem [{\citenamefont {Tourin}\ \emph {et~al.}(1997)\citenamefont {Tourin},
  \citenamefont {Derode}, \citenamefont {Roux}, \citenamefont {van Tiggelen},\
  and\ \citenamefont {Fink}}]{Tourin1997}%
  \BibitemOpen
  \bibfield  {author} {\bibinfo {author} {\bibfnamefont {Arnaud}\ \bibnamefont
  {Tourin}}, \bibinfo {author} {\bibfnamefont {Arnaud}\ \bibnamefont {Derode}},
  \bibinfo {author} {\bibfnamefont {Philippe}\ \bibnamefont {Roux}}, \bibinfo
  {author} {\bibfnamefont {Bart~A.}\ \bibnamefont {van Tiggelen}}, \ and\
  \bibinfo {author} {\bibfnamefont {Mathias}\ \bibnamefont {Fink}},\ }\bibfield
   {title} {\enquote {\bibinfo {title} {Time-dependent coherent backscattering
  of acoustic waves},}\ }\href {\doibase 10.1103/physrevlett.79.3637}
  {\bibfield  {journal} {\bibinfo  {journal} {Physical Review Letters}\
  }\textbf {\bibinfo {volume} {79}},\ \bibinfo {pages} {3637--3639} (\bibinfo
  {year} {1997})}\BibitemShut {NoStop}%
\bibitem [{\citenamefont {Larose}\ \emph {et~al.}(2004)\citenamefont {Larose},
  \citenamefont {Margerin}, \citenamefont {van Tiggelen},\ and\ \citenamefont
  {Campillo}}]{Larose2004}%
  \BibitemOpen
  \bibfield  {author} {\bibinfo {author} {\bibfnamefont {E.}~\bibnamefont
  {Larose}}, \bibinfo {author} {\bibfnamefont {L.}~\bibnamefont {Margerin}},
  \bibinfo {author} {\bibfnamefont {B.~A.}\ \bibnamefont {van Tiggelen}}, \
  and\ \bibinfo {author} {\bibfnamefont {M.}~\bibnamefont {Campillo}},\
  }\bibfield  {title} {\enquote {\bibinfo {title} {Weak localization of seismic
  waves},}\ }\href {\doibase 10.1103/physrevlett.93.048501} {\bibfield
  {journal} {\bibinfo  {journal} {Physical Review Letters}\ }\textbf {\bibinfo
  {volume} {93}} (\bibinfo {year} {2004}),\
  10.1103/physrevlett.93.048501}\BibitemShut {NoStop}%
\bibitem [{\citenamefont {Jendrzejewski}\ \emph {et~al.}(2012)\citenamefont
  {Jendrzejewski}, \citenamefont {M\"{u}ller}, \citenamefont {Richard},
  \citenamefont {Date}, \citenamefont {Plisson}, \citenamefont {Bouyer},
  \citenamefont {Aspect},\ and\ \citenamefont {Josse}}]{Jendrzejewski2012}%
  \BibitemOpen
  \bibfield  {author} {\bibinfo {author} {\bibfnamefont {F.}~\bibnamefont
  {Jendrzejewski}}, \bibinfo {author} {\bibfnamefont {K.}~\bibnamefont
  {M\"{u}ller}}, \bibinfo {author} {\bibfnamefont {J.}~\bibnamefont {Richard}},
  \bibinfo {author} {\bibfnamefont {A.}~\bibnamefont {Date}}, \bibinfo {author}
  {\bibfnamefont {T.}~\bibnamefont {Plisson}}, \bibinfo {author} {\bibfnamefont
  {P.}~\bibnamefont {Bouyer}}, \bibinfo {author} {\bibfnamefont
  {A.}~\bibnamefont {Aspect}}, \ and\ \bibinfo {author} {\bibfnamefont
  {V.}~\bibnamefont {Josse}},\ }\bibfield  {title} {\enquote {\bibinfo {title}
  {Coherent backscattering of ultracold atoms},}\ }\href {\doibase
  10.1103/physrevlett.109.195302} {\bibfield  {journal} {\bibinfo  {journal}
  {Physical Review Letters}\ }\textbf {\bibinfo {volume} {109}} (\bibinfo
  {year} {2012}),\ 10.1103/physrevlett.109.195302}\BibitemShut {NoStop}%
\bibitem [{\citenamefont {Labeyrie}\ \emph {et~al.}(1999)\citenamefont
  {Labeyrie}, \citenamefont {de~Tomasi}, \citenamefont {Bernard}, \citenamefont
  {M\"{u}ller}, \citenamefont {Miniatura},\ and\ \citenamefont
  {Kaiser}}]{Labeyrie1999}%
  \BibitemOpen
  \bibfield  {author} {\bibinfo {author} {\bibfnamefont {G.}~\bibnamefont
  {Labeyrie}}, \bibinfo {author} {\bibfnamefont {F.}~\bibnamefont {de~Tomasi}},
  \bibinfo {author} {\bibfnamefont {J.-C.}\ \bibnamefont {Bernard}}, \bibinfo
  {author} {\bibfnamefont {C.~A.}\ \bibnamefont {M\"{u}ller}}, \bibinfo
  {author} {\bibfnamefont {C.}~\bibnamefont {Miniatura}}, \ and\ \bibinfo
  {author} {\bibfnamefont {R.}~\bibnamefont {Kaiser}},\ }\bibfield  {title}
  {\enquote {\bibinfo {title} {Coherent backscattering of light by cold
  atoms},}\ }\href {\doibase 10.1103/physrevlett.83.5266} {\bibfield  {journal}
  {\bibinfo  {journal} {Physical Review Letters}\ }\textbf {\bibinfo {volume}
  {83}},\ \bibinfo {pages} {5266--5269} (\bibinfo {year} {1999})}\BibitemShut
  {NoStop}%
\bibitem [{\citenamefont {Jonckheere}\ \emph {et~al.}(2000)\citenamefont
  {Jonckheere}, \citenamefont {M\"{u}ller}, \citenamefont {Kaiser},
  \citenamefont {Miniatura},\ and\ \citenamefont {Delande}}]{Jonckheere2000}%
  \BibitemOpen
  \bibfield  {author} {\bibinfo {author} {\bibfnamefont {Thibaut}\ \bibnamefont
  {Jonckheere}}, \bibinfo {author} {\bibfnamefont {Cord~A.}\ \bibnamefont
  {M\"{u}ller}}, \bibinfo {author} {\bibfnamefont {Robin}\ \bibnamefont
  {Kaiser}}, \bibinfo {author} {\bibfnamefont {Christian}\ \bibnamefont
  {Miniatura}}, \ and\ \bibinfo {author} {\bibfnamefont {Dominique}\
  \bibnamefont {Delande}},\ }\bibfield  {title} {\enquote {\bibinfo {title}
  {Multiple scattering of light by atoms in the weak localization regime},}\
  }\href {\doibase 10.1103/physrevlett.85.4269} {\bibfield  {journal} {\bibinfo
   {journal} {Physical Review Letters}\ }\textbf {\bibinfo {volume} {85}},\
  \bibinfo {pages} {4269--4272} (\bibinfo {year} {2000})}\BibitemShut {NoStop}%
\bibitem [{\citenamefont {Bidel}\ \emph {et~al.}(2002)\citenamefont {Bidel},
  \citenamefont {Klappauf}, \citenamefont {Bernard}, \citenamefont {Delande},
  \citenamefont {Labeyrie}, \citenamefont {Miniatura}, \citenamefont
  {Wilkowski},\ and\ \citenamefont {Kaiser}}]{Bidel2002}%
  \BibitemOpen
  \bibfield  {author} {\bibinfo {author} {\bibfnamefont {Y.}~\bibnamefont
  {Bidel}}, \bibinfo {author} {\bibfnamefont {B.}~\bibnamefont {Klappauf}},
  \bibinfo {author} {\bibfnamefont {J.~C.}\ \bibnamefont {Bernard}}, \bibinfo
  {author} {\bibfnamefont {D.}~\bibnamefont {Delande}}, \bibinfo {author}
  {\bibfnamefont {G.}~\bibnamefont {Labeyrie}}, \bibinfo {author}
  {\bibfnamefont {C.}~\bibnamefont {Miniatura}}, \bibinfo {author}
  {\bibfnamefont {D.}~\bibnamefont {Wilkowski}}, \ and\ \bibinfo {author}
  {\bibfnamefont {R.}~\bibnamefont {Kaiser}},\ }\bibfield  {title} {\enquote
  {\bibinfo {title} {Coherent light transport in a cold strontium cloud},}\
  }\href {\doibase 10.1103/physrevlett.88.203902} {\bibfield  {journal}
  {\bibinfo  {journal} {Physical Review Letters}\ }\textbf {\bibinfo {volume}
  {88}} (\bibinfo {year} {2002}),\ 10.1103/physrevlett.88.203902}\BibitemShut
  {NoStop}%
\bibitem [{\citenamefont {Kupriyanov}\ \emph {et~al.}(2003)\citenamefont
  {Kupriyanov}, \citenamefont {Sokolov}, \citenamefont {Kulatunga},
  \citenamefont {Sukenik},\ and\ \citenamefont {Havey}}]{Kupriyanov2003}%
  \BibitemOpen
  \bibfield  {author} {\bibinfo {author} {\bibfnamefont {D.~V.}\ \bibnamefont
  {Kupriyanov}}, \bibinfo {author} {\bibfnamefont {I.~M.}\ \bibnamefont
  {Sokolov}}, \bibinfo {author} {\bibfnamefont {P.}~\bibnamefont {Kulatunga}},
  \bibinfo {author} {\bibfnamefont {C.~I.}\ \bibnamefont {Sukenik}}, \ and\
  \bibinfo {author} {\bibfnamefont {M.~D.}\ \bibnamefont {Havey}},\ }\bibfield
  {title} {\enquote {\bibinfo {title} {Coherent backscattering of light in
  atomic systems: Application to weak localization in an ensemble of cold
  alkali-metal atoms},}\ }\href {\doibase 10.1103/physreva.67.013814}
  {\bibfield  {journal} {\bibinfo  {journal} {Physical Review A}\ }\textbf
  {\bibinfo {volume} {67}} (\bibinfo {year} {2003}),\
  10.1103/physreva.67.013814}\BibitemShut {NoStop}%
\bibitem [{\citenamefont {T.}\ \emph {et~al.}(2004)\citenamefont {T.},
  \citenamefont {D.}, \citenamefont {Y.}, \citenamefont {R.},\ and\
  \citenamefont {C.}}]{Chaneliere2004}%
  \BibitemOpen
  \bibfield  {author} {\bibinfo {author} {\bibfnamefont {Chaneli\`ere}\
  \bibnamefont {T.}}, \bibinfo {author} {\bibfnamefont {Wilkowski}\
  \bibnamefont {D.}}, \bibinfo {author} {\bibfnamefont {Bidel}\ \bibnamefont
  {Y.}}, \bibinfo {author} {\bibfnamefont {Kaiser}\ \bibnamefont {R.}}, \ and\
  \bibinfo {author} {\bibfnamefont {Miniatura}\ \bibnamefont {C.}},\ }\bibfield
   {title} {\enquote {\bibinfo {title} {Saturation-induced coherence loss in
  coherent backscattering of light},}\ }\href@noop {} {\bibfield  {journal}
  {\bibinfo  {journal} {Phys. Rev. E}\ }\textbf {\bibinfo {volume} {70}},\
  \bibinfo {pages} {036602} (\bibinfo {year} {2004})}\BibitemShut {NoStop}%
\bibitem [{\citenamefont {Kupriyanov}\ \emph {et~al.}(2006)\citenamefont
  {Kupriyanov}, \citenamefont {Sokolov}, \citenamefont {Sukenik},\ and\
  \citenamefont {Havey}}]{Kupriyanov2006}%
  \BibitemOpen
  \bibfield  {author} {\bibinfo {author} {\bibfnamefont {D~V}\ \bibnamefont
  {Kupriyanov}}, \bibinfo {author} {\bibfnamefont {I~M}\ \bibnamefont
  {Sokolov}}, \bibinfo {author} {\bibfnamefont {C~I}\ \bibnamefont {Sukenik}},
  \ and\ \bibinfo {author} {\bibfnamefont {M~D}\ \bibnamefont {Havey}},\
  }\bibfield  {title} {\enquote {\bibinfo {title} {Coherent backscattering of
  light from ultracold and optically dense atomic ensembles},}\ }\href
  {\doibase 10.1002/lapl.200510059} {\bibfield  {journal} {\bibinfo  {journal}
  {Laser Physics Letters}\ }\textbf {\bibinfo {volume} {3}},\ \bibinfo {pages}
  {223--243} (\bibinfo {year} {2006})}\BibitemShut {NoStop}%
\bibitem [{\citenamefont {Labeyrie}(2008)}]{Labeyrie2008}%
  \BibitemOpen
  \bibfield  {author} {\bibinfo {author} {\bibfnamefont {G.}~\bibnamefont
  {Labeyrie}},\ }\bibfield  {title} {\enquote {\bibinfo {title} {Coherent
  transport of light in cold atoms},}\ }\href {\doibase
  10.1142/s0217984908014699} {\bibfield  {journal} {\bibinfo  {journal} {Modern
  Physics Letters B}\ }\textbf {\bibinfo {volume} {22}},\ \bibinfo {pages}
  {73--99} (\bibinfo {year} {2008})}\BibitemShut {NoStop}%
\bibitem [{\citenamefont {Wellens}\ \emph {et~al.}(2004)\citenamefont
  {Wellens}, \citenamefont {Gr{\'{e}}maud}, \citenamefont {Delande},\ and\
  \citenamefont {Miniatura}}]{Wellens2004}%
  \BibitemOpen
  \bibfield  {author} {\bibinfo {author} {\bibfnamefont {T.}~\bibnamefont
  {Wellens}}, \bibinfo {author} {\bibfnamefont {B.}~\bibnamefont
  {Gr{\'{e}}maud}}, \bibinfo {author} {\bibfnamefont {D.}~\bibnamefont
  {Delande}}, \ and\ \bibinfo {author} {\bibfnamefont {C.}~\bibnamefont
  {Miniatura}},\ }\bibfield  {title} {\enquote {\bibinfo {title} {Coherent
  backscattering of light by two atoms in the saturated regime},}\ }\href
  {\doibase 10.1103/physreva.70.023817} {\bibfield  {journal} {\bibinfo
  {journal} {Physical Review A}\ }\textbf {\bibinfo {volume} {70}} (\bibinfo
  {year} {2004}),\ 10.1103/physreva.70.023817}\BibitemShut {NoStop}%
\bibitem [{\citenamefont {Totsuka}\ and\ \citenamefont
  {Tomita}(1999)}]{Totsuka1999}%
  \BibitemOpen
  \bibfield  {author} {\bibinfo {author} {\bibfnamefont {Kouki}\ \bibnamefont
  {Totsuka}}\ and\ \bibinfo {author} {\bibfnamefont {Makoto}\ \bibnamefont
  {Tomita}},\ }\bibfield  {title} {\enquote {\bibinfo {title} {Coherent
  backscattering in a disordered optical medium in the presence of saturation
  absorption},}\ }\href {\doibase 10.1103/physrevb.59.11139} {\bibfield
  {journal} {\bibinfo  {journal} {Physical Review B}\ }\textbf {\bibinfo
  {volume} {59}},\ \bibinfo {pages} {11139--11142} (\bibinfo {year}
  {1999})}\BibitemShut {NoStop}%
\bibitem [{\citenamefont {Balik}\ \emph {et~al.}(2005)\citenamefont {Balik},
  \citenamefont {Kulatunga}, \citenamefont {Sukenik}, \citenamefont {Havey},
  \citenamefont {Kupriyanov},\ and\ \citenamefont {Sokolov}}]{Balik2005}%
  \BibitemOpen
  \bibfield  {author} {\bibinfo {author} {\bibfnamefont {S.}~\bibnamefont
  {Balik}}, \bibinfo {author} {\bibfnamefont {P.}~\bibnamefont {Kulatunga}},
  \bibinfo {author} {\bibfnamefont {C.~I.}\ \bibnamefont {Sukenik}}, \bibinfo
  {author} {\bibfnamefont {M.~D.}\ \bibnamefont {Havey}}, \bibinfo {author}
  {\bibfnamefont {D.~V.}\ \bibnamefont {Kupriyanov}}, \ and\ \bibinfo {author}
  {\bibfnamefont {I.~M.}\ \bibnamefont {Sokolov}},\ }\bibfield  {title}
  {\enquote {\bibinfo {title} {Strong-field coherent backscattering of light in
  ultracold atomic85rb},}\ }\href {\doibase 10.1080/09500340500275934}
  {\bibfield  {journal} {\bibinfo  {journal} {Journal of Modern Optics}\
  }\textbf {\bibinfo {volume} {52}},\ \bibinfo {pages} {2269--2278} (\bibinfo
  {year} {2005})}\BibitemShut {NoStop}%
\bibitem [{\citenamefont {Shatokhin}\ \emph {et~al.}(2005)\citenamefont
  {Shatokhin}, \citenamefont {M\"{u}ller},\ and\ \citenamefont
  {Buchleitner}}]{Shatokhin2005}%
  \BibitemOpen
  \bibfield  {author} {\bibinfo {author} {\bibfnamefont {V.}~\bibnamefont
  {Shatokhin}}, \bibinfo {author} {\bibfnamefont {C.~A.}\ \bibnamefont
  {M\"{u}ller}}, \ and\ \bibinfo {author} {\bibfnamefont {A.}~\bibnamefont
  {Buchleitner}},\ }\bibfield  {title} {\enquote {\bibinfo {title} {Coherent
  inelastic backscattering of intense laser light by cold atoms},}\ }\href
  {\doibase 10.1103/physrevlett.94.043603} {\bibfield  {journal} {\bibinfo
  {journal} {Physical Review Letters}\ }\textbf {\bibinfo {volume} {94}}
  (\bibinfo {year} {2005}),\ 10.1103/physrevlett.94.043603}\BibitemShut
  {NoStop}%
\bibitem [{\citenamefont {Greffet}(1991)}]{Greffet1991}%
  \BibitemOpen
  \bibfield  {author} {\bibinfo {author} {\bibfnamefont {Jean-Jacques}\
  \bibnamefont {Greffet}},\ }\bibfield  {title} {\enquote {\bibinfo {title}
  {Backscattering of s-polarized light from a cloud of small particles above a
  dielectric substrate},}\ }\href {\doibase 10.1088/0959-7174/1/3/006}
  {\bibfield  {journal} {\bibinfo  {journal} {Waves in Random Media}\ }\textbf
  {\bibinfo {volume} {1}},\ \bibinfo {pages} {S65--S73} (\bibinfo {year}
  {1991})}\BibitemShut {NoStop}%
\bibitem [{\citenamefont {Labeyrie}\ \emph {et~al.}(2000)\citenamefont
  {Labeyrie}, \citenamefont {M\"{u}ller}, \citenamefont {Wiersma},
  \citenamefont {Miniatura},\ and\ \citenamefont {Kaiser}}]{Labeyrie2000}%
  \BibitemOpen
  \bibfield  {author} {\bibinfo {author} {\bibfnamefont {G}~\bibnamefont
  {Labeyrie}}, \bibinfo {author} {\bibfnamefont {C~A}\ \bibnamefont
  {M\"{u}ller}}, \bibinfo {author} {\bibfnamefont {D~S}\ \bibnamefont
  {Wiersma}}, \bibinfo {author} {\bibfnamefont {Ch}~\bibnamefont {Miniatura}},
  \ and\ \bibinfo {author} {\bibfnamefont {R}~\bibnamefont {Kaiser}},\
  }\bibfield  {title} {\enquote {\bibinfo {title} {Observation of coherent
  backscattering of light by cold atoms},}\ }\href {\doibase
  10.1088/1464-4266/2/5/316} {\bibfield  {journal} {\bibinfo  {journal}
  {Journal of Optics B: Quantum and Semiclassical Optics}\ }\textbf {\bibinfo
  {volume} {2}},\ \bibinfo {pages} {672--685} (\bibinfo {year}
  {2000})}\BibitemShut {NoStop}%
\bibitem [{\citenamefont {Moriya}\ \emph {et~al.}(2016)\citenamefont {Moriya},
  \citenamefont {Shiozaki}, \citenamefont {Teixeira}, \citenamefont
  {M{\'{a}}ximo}, \citenamefont {Piovella}, \citenamefont {Bachelard},
  \citenamefont {Kaiser},\ and\ \citenamefont {Courteille}}]{Moriya2016}%
  \BibitemOpen
  \bibfield  {author} {\bibinfo {author} {\bibfnamefont {P.~H.}\ \bibnamefont
  {Moriya}}, \bibinfo {author} {\bibfnamefont {R.~F.}\ \bibnamefont
  {Shiozaki}}, \bibinfo {author} {\bibfnamefont {R.~Celistrino}\ \bibnamefont
  {Teixeira}}, \bibinfo {author} {\bibfnamefont {C.~E.}\ \bibnamefont
  {M{\'{a}}ximo}}, \bibinfo {author} {\bibfnamefont {N.}~\bibnamefont
  {Piovella}}, \bibinfo {author} {\bibfnamefont {R.}~\bibnamefont {Bachelard}},
  \bibinfo {author} {\bibfnamefont {R.}~\bibnamefont {Kaiser}}, \ and\ \bibinfo
  {author} {\bibfnamefont {Ph.~W.}\ \bibnamefont {Courteille}},\ }\bibfield
  {title} {\enquote {\bibinfo {title} {Coherent backscattering of inelastic
  photons from atoms and their mirror images},}\ }\href {\doibase
  10.1103/physreva.94.053806} {\bibfield  {journal} {\bibinfo  {journal}
  {Physical Review A}\ }\textbf {\bibinfo {volume} {94}} (\bibinfo {year}
  {2016}),\ 10.1103/physreva.94.053806}\BibitemShut {NoStop}%
\bibitem [{\citenamefont {Mollow}(1969)}]{Mollow1969}%
  \BibitemOpen
  \bibfield  {author} {\bibinfo {author} {\bibfnamefont {B.~R.}\ \bibnamefont
  {Mollow}},\ }\bibfield  {title} {\enquote {\bibinfo {title} {Power spectrum
  of light scattered by two-level systems},}\ }\href {\doibase
  10.1103/physrev.188.1969} {\bibfield  {journal} {\bibinfo  {journal}
  {Physical Review}\ }\textbf {\bibinfo {volume} {188}},\ \bibinfo {pages}
  {1969--1975} (\bibinfo {year} {1969})}\BibitemShut {NoStop}%
\bibitem [{\citenamefont {Schuda}\ \emph {et~al.}(1974)\citenamefont {Schuda},
  \citenamefont {Jr.},\ and\ \citenamefont {Hercher}}]{Schuda1974}%
  \BibitemOpen
  \bibfield  {author} {\bibinfo {author} {\bibfnamefont {F.}~\bibnamefont
  {Schuda}}, \bibinfo {author} {\bibfnamefont {C.~R.~Stroud}\ \bibnamefont
  {Jr.}}, \ and\ \bibinfo {author} {\bibfnamefont {M.}~\bibnamefont
  {Hercher}},\ }\bibfield  {title} {\enquote {\bibinfo {title} {Observation of
  the resonant stark effect at optical frequencies},}\ }\href@noop {}
  {\bibfield  {journal} {\bibinfo  {journal} {J. Phys. B: Atom. Molec. Phys.}\
  }\textbf {\bibinfo {volume} {7}},\ \bibinfo {pages} {L198} (\bibinfo {year}
  {1974})}\BibitemShut {NoStop}%
\bibitem [{\citenamefont {Wu}\ \emph {et~al.}(1975)\citenamefont {Wu},
  \citenamefont {Grove},\ and\ \citenamefont {Ezekiel}}]{Wu1975}%
  \BibitemOpen
  \bibfield  {author} {\bibinfo {author} {\bibfnamefont {F.~Y.}\ \bibnamefont
  {Wu}}, \bibinfo {author} {\bibfnamefont {R.~E.}\ \bibnamefont {Grove}}, \
  and\ \bibinfo {author} {\bibfnamefont {S.}~\bibnamefont {Ezekiel}},\
  }\bibfield  {title} {\enquote {\bibinfo {title} {Investigation of the
  spectrum of resonance fluorescence induced by a monochromatic field},}\
  }\href@noop {} {\bibfield  {journal} {\bibinfo  {journal} {Phys. Rev. Lett.}\
  }\textbf {\bibinfo {volume} {35}},\ \bibinfo {pages} {1426} (\bibinfo {year}
  {1975})}\BibitemShut {NoStop}%
\bibitem [{\citenamefont {Stalgies}\ \emph {et~al.}(1996)\citenamefont
  {Stalgies}, \citenamefont {Siemers}, \citenamefont {Appasamy}, \citenamefont
  {Altevogt},\ and\ \citenamefont {Toschek}}]{Stalgies1996}%
  \BibitemOpen
  \bibfield  {author} {\bibinfo {author} {\bibfnamefont {Y}~\bibnamefont
  {Stalgies}}, \bibinfo {author} {\bibfnamefont {I}~\bibnamefont {Siemers}},
  \bibinfo {author} {\bibfnamefont {B}~\bibnamefont {Appasamy}}, \bibinfo
  {author} {\bibfnamefont {T}~\bibnamefont {Altevogt}}, \ and\ \bibinfo
  {author} {\bibfnamefont {P.~E}\ \bibnamefont {Toschek}},\ }\bibfield  {title}
  {\enquote {\bibinfo {title} {The spectrum of single-atom resonance
  fluorescence},}\ }\href {\doibase 10.1209/epl/i1996-00563-6} {\bibfield
  {journal} {\bibinfo  {journal} {Europhysics Letters ({EPL})}\ }\textbf
  {\bibinfo {volume} {35}},\ \bibinfo {pages} {259--264} (\bibinfo {year}
  {1996})}\BibitemShut {NoStop}%
\bibitem [{\citenamefont {Wrigge}\ \emph {et~al.}(2007)\citenamefont {Wrigge},
  \citenamefont {Gerhardt}, \citenamefont {Hwang}, \citenamefont {Zumofen},\
  and\ \citenamefont {Sandoghdar}}]{Wrigge2007}%
  \BibitemOpen
  \bibfield  {author} {\bibinfo {author} {\bibfnamefont {G.}~\bibnamefont
  {Wrigge}}, \bibinfo {author} {\bibfnamefont {I.}~\bibnamefont {Gerhardt}},
  \bibinfo {author} {\bibfnamefont {J.}~\bibnamefont {Hwang}}, \bibinfo
  {author} {\bibfnamefont {G.}~\bibnamefont {Zumofen}}, \ and\ \bibinfo
  {author} {\bibfnamefont {V.}~\bibnamefont {Sandoghdar}},\ }\bibfield  {title}
  {\enquote {\bibinfo {title} {Efficient coupling of photons to a single
  molecule and the observation of its resonance fluorescence},}\ }\href@noop {}
  {\bibfield  {journal} {\bibinfo  {journal} {Nat. Phys.}\ }\textbf {\bibinfo
  {volume} {4}},\ \bibinfo {pages} {60} (\bibinfo {year} {2007})}\BibitemShut
  {NoStop}%
\bibitem [{\citenamefont {Muller}\ \emph {et~al.}(2007)\citenamefont {Muller},
  \citenamefont {anbd P.~Bianucci}, \citenamefont {Wang}, \citenamefont
  {Deppe}, \citenamefont {Ma}, \citenamefont {Zhang}, \citenamefont {Salamo},
  \citenamefont {Xiao},\ and\ \citenamefont {Shih}}]{Muller2007}%
  \BibitemOpen
  \bibfield  {author} {\bibinfo {author} {\bibfnamefont {A.}~\bibnamefont
  {Muller}}, \bibinfo {author} {\bibfnamefont {E.~B.~Flagg}\ \bibnamefont {anbd
  P.~Bianucci}}, \bibinfo {author} {\bibfnamefont {X.~Y.}\ \bibnamefont
  {Wang}}, \bibinfo {author} {\bibfnamefont {D.~G.}\ \bibnamefont {Deppe}},
  \bibinfo {author} {\bibfnamefont {W.}~\bibnamefont {Ma}}, \bibinfo {author}
  {\bibfnamefont {J.}~\bibnamefont {Zhang}}, \bibinfo {author} {\bibfnamefont
  {G.~J.}\ \bibnamefont {Salamo}}, \bibinfo {author} {\bibfnamefont
  {M.}~\bibnamefont {Xiao}}, \ and\ \bibinfo {author} {\bibfnamefont {C.~K.}\
  \bibnamefont {Shih}},\ }\bibfield  {title} {\enquote {\bibinfo {title}
  {Resonance fluorescence from a coherently driven semiconductor quantum dot in
  a cavity},}\ }\href@noop {} {\bibfield  {journal} {\bibinfo  {journal} {Phys.
  Rev. Lett.}\ }\textbf {\bibinfo {volume} {99}},\ \bibinfo {pages} {187402}
  (\bibinfo {year} {2007})}\BibitemShut {NoStop}%
\bibitem [{\citenamefont {Vamivakas}\ \emph {et~al.}(2009)\citenamefont
  {Vamivakas}, \citenamefont {Zhao}, \citenamefont {Lu},\ and\ \citenamefont
  {Atat\"ure}}]{Vamivakas2009}%
  \BibitemOpen
  \bibfield  {author} {\bibinfo {author} {\bibfnamefont {A.~Nick}\ \bibnamefont
  {Vamivakas}}, \bibinfo {author} {\bibfnamefont {Y.}~\bibnamefont {Zhao}},
  \bibinfo {author} {\bibfnamefont {C.-Y.}\ \bibnamefont {Lu}}, \ and\ \bibinfo
  {author} {\bibfnamefont {M.}~\bibnamefont {Atat\"ure}},\ }\bibfield  {title}
  {\enquote {\bibinfo {title} {Spin-resolved quantuum-dot resonance
  fluorescence},}\ }\href@noop {} {\bibfield  {journal} {\bibinfo  {journal}
  {Nat. Phys.}\ }\textbf {\bibinfo {volume} {5}},\ \bibinfo {pages} {198}
  (\bibinfo {year} {2009})}\BibitemShut {NoStop}%
\bibitem [{\citenamefont {Flagg}\ \emph {et~al.}(2009)\citenamefont {Flagg},
  \citenamefont {Muller}, \citenamefont {Robertson}, \citenamefont {Founta},
  \citenamefont {Deppe}, \citenamefont {Xiao}, \citenamefont {Ma},
  \citenamefont {Salamo},\ and\ \citenamefont {Shih}}]{Flagg2009}%
  \BibitemOpen
  \bibfield  {author} {\bibinfo {author} {\bibfnamefont {E.~B.}\ \bibnamefont
  {Flagg}}, \bibinfo {author} {\bibfnamefont {A.}~\bibnamefont {Muller}},
  \bibinfo {author} {\bibfnamefont {J.W.}\ \bibnamefont {Robertson}}, \bibinfo
  {author} {\bibfnamefont {S.}~\bibnamefont {Founta}}, \bibinfo {author}
  {\bibfnamefont {D.~G.}\ \bibnamefont {Deppe}}, \bibinfo {author}
  {\bibfnamefont {M.}~\bibnamefont {Xiao}}, \bibinfo {author} {\bibfnamefont
  {W.}~\bibnamefont {Ma}}, \bibinfo {author} {\bibfnamefont {G.~J.}\
  \bibnamefont {Salamo}}, \ and\ \bibinfo {author} {\bibfnamefont {C.~K.}\
  \bibnamefont {Shih}},\ }\bibfield  {title} {\enquote {\bibinfo {title}
  {Resonantly driven coherent oscillations in a solid-state quantum emitter},}\
  }\href@noop {} {\bibfield  {journal} {\bibinfo  {journal} {Nat. Phys.}\
  }\textbf {\bibinfo {volume} {5}},\ \bibinfo {pages} {203} (\bibinfo {year}
  {2009})}\BibitemShut {NoStop}%
\bibitem [{\citenamefont {Zhou}\ \emph {et~al.}(2017)\citenamefont {Zhou},
  \citenamefont {Rasmita}, \citenamefont {Li}, \citenamefont {Xiong},
  \citenamefont {Aharonovich},\ and\ \citenamefont {Gao}}]{Zhou2017}%
  \BibitemOpen
  \bibfield  {author} {\bibinfo {author} {\bibfnamefont {Y.}~\bibnamefont
  {Zhou}}, \bibinfo {author} {\bibfnamefont {A.}~\bibnamefont {Rasmita}},
  \bibinfo {author} {\bibfnamefont {K.}~\bibnamefont {Li}}, \bibinfo {author}
  {\bibfnamefont {Q.}~\bibnamefont {Xiong}}, \bibinfo {author} {\bibfnamefont
  {I.}~\bibnamefont {Aharonovich}}, \ and\ \bibinfo {author} {\bibfnamefont
  {W.-B.}\ \bibnamefont {Gao}},\ }\bibfield  {title} {\enquote {\bibinfo
  {title} {Coherent control of a strongly driven silicon vacancy optical
  transition in diamond},}\ }\href@noop {} {\bibfield  {journal} {\bibinfo
  {journal} {Nat. Comm.}\ }\textbf {\bibinfo {volume} {8}},\ \bibinfo {pages}
  {14451} (\bibinfo {year} {2017})}\BibitemShut {NoStop}%
\bibitem [{\citenamefont {Saiko}\ \emph {et~al.}(2014)\citenamefont {Saiko},
  \citenamefont {Fedaruk},\ and\ \citenamefont {Markevich}}]{Saiko2014}%
  \BibitemOpen
  \bibfield  {author} {\bibinfo {author} {\bibfnamefont {A.~P.}\ \bibnamefont
  {Saiko}}, \bibinfo {author} {\bibfnamefont {R.}~\bibnamefont {Fedaruk}}, \
  and\ \bibinfo {author} {\bibfnamefont {S.~A.}\ \bibnamefont {Markevich}},\
  }\bibfield  {title} {\enquote {\bibinfo {title} {Detuning dependent narrowing
  of mollow triplet lines of driven quantum dots},}\ }\href@noop {} {\bibfield
  {journal} {\bibinfo  {journal} {J. Exp. Theor. Phys.}\ }\textbf {\bibinfo
  {volume} {118}},\ \bibinfo {pages} {655} (\bibinfo {year}
  {2014})}\BibitemShut {NoStop}%
\bibitem [{\citenamefont {Toyli}\ \emph {et~al.}(2016)\citenamefont {Toyli},
  \citenamefont {Eddins}, \citenamefont {Boutin}, \citenamefont {Puri},
  \citenamefont {Hover}, \citenamefont {Bolkhovsky}, \citenamefont {Oliver},
  \citenamefont {Blais},\ and\ \citenamefont {Siddiqi}}]{Toyli2016}%
  \BibitemOpen
  \bibfield  {author} {\bibinfo {author} {\bibfnamefont {D.~M.}\ \bibnamefont
  {Toyli}}, \bibinfo {author} {\bibfnamefont {A.W.}\ \bibnamefont {Eddins}},
  \bibinfo {author} {\bibfnamefont {S.}~\bibnamefont {Boutin}}, \bibinfo
  {author} {\bibfnamefont {S.}~\bibnamefont {Puri}}, \bibinfo {author}
  {\bibfnamefont {D.}~\bibnamefont {Hover}}, \bibinfo {author} {\bibfnamefont
  {V.}~\bibnamefont {Bolkhovsky}}, \bibinfo {author} {\bibfnamefont {W.~D.}\
  \bibnamefont {Oliver}}, \bibinfo {author} {\bibfnamefont {A.}~\bibnamefont
  {Blais}}, \ and\ \bibinfo {author} {\bibfnamefont {I.}~\bibnamefont
  {Siddiqi}},\ }\bibfield  {title} {\enquote {\bibinfo {title} {Resonance
  fluorescence from an artificial atom in squeezed vacuum},}\ }\href@noop {}
  {\bibfield  {journal} {\bibinfo  {journal} {Phys. Rev. X}\ }\textbf {\bibinfo
  {volume} {6}},\ \bibinfo {pages} {031004} (\bibinfo {year}
  {2016})}\BibitemShut {NoStop}%
\bibitem [{\citenamefont {Kim}\ \emph {et~al.}(2014)\citenamefont {Kim},
  \citenamefont {Shen}, \citenamefont {Roy-Choudhury}, \citenamefont
  {Solomon},\ and\ \citenamefont {Waks}}]{Kim2014}%
  \BibitemOpen
  \bibfield  {author} {\bibinfo {author} {\bibfnamefont {H.}~\bibnamefont
  {Kim}}, \bibinfo {author} {\bibfnamefont {T.~C.}\ \bibnamefont {Shen}},
  \bibinfo {author} {\bibfnamefont {K.}~\bibnamefont {Roy-Choudhury}}, \bibinfo
  {author} {\bibfnamefont {G.~S.}\ \bibnamefont {Solomon}}, \ and\ \bibinfo
  {author} {\bibfnamefont {E.}~\bibnamefont {Waks}},\ }\bibfield  {title}
  {\enquote {\bibinfo {title} {Resonant interactions between amollow triplet
  sideband and a strongly coupled cavity},}\ }\href@noop {} {\bibfield
  {journal} {\bibinfo  {journal} {Phys. Rev. Lett.}\ }\textbf {\bibinfo
  {volume} {114}},\ \bibinfo {pages} {027403} (\bibinfo {year}
  {2014})}\BibitemShut {NoStop}%
\bibitem [{\citenamefont {Ott}\ \emph {et~al.}(2013)\citenamefont {Ott},
  \citenamefont {Wubs}, \citenamefont {Lodahl}, \citenamefont {Mortensen},\
  and\ \citenamefont {Kaiser}}]{Ott2013}%
  \BibitemOpen
  \bibfield  {author} {\bibinfo {author} {\bibfnamefont {J.~R.}\ \bibnamefont
  {Ott}}, \bibinfo {author} {\bibfnamefont {M.}~\bibnamefont {Wubs}}, \bibinfo
  {author} {\bibfnamefont {P.}~\bibnamefont {Lodahl}}, \bibinfo {author}
  {\bibfnamefont {N.~A.}\ \bibnamefont {Mortensen}}, \ and\ \bibinfo {author}
  {\bibfnamefont {R.}~\bibnamefont {Kaiser}},\ }\bibfield  {title} {\enquote
  {\bibinfo {title} {Cooperative fluorescence from a strongly driven dilute
  cloud of atoms},}\ }\href {\doibase 10.1103/physreva.87.061801} {\bibfield
  {journal} {\bibinfo  {journal} {Physical Review A}\ }\textbf {\bibinfo
  {volume} {87}} (\bibinfo {year} {2013}),\
  10.1103/physreva.87.061801}\BibitemShut {NoStop}%
\bibitem [{\citenamefont {Pucci}\ \emph {et~al.}(2017)\citenamefont {Pucci},
  \citenamefont {Roy}, \citenamefont {do~Espirito~Santo}, \citenamefont
  {Kaiser}, \citenamefont {Kastner},\ and\ \citenamefont
  {Bachelard}}]{Pucci2017}%
  \BibitemOpen
  \bibfield  {author} {\bibinfo {author} {\bibfnamefont {Lorenzo}\ \bibnamefont
  {Pucci}}, \bibinfo {author} {\bibfnamefont {Analabha}\ \bibnamefont {Roy}},
  \bibinfo {author} {\bibfnamefont {Tiago~Santiago}\ \bibnamefont
  {do~Espirito~Santo}}, \bibinfo {author} {\bibfnamefont {Robin}\ \bibnamefont
  {Kaiser}}, \bibinfo {author} {\bibfnamefont {Michael}\ \bibnamefont
  {Kastner}}, \ and\ \bibinfo {author} {\bibfnamefont {Romain}\ \bibnamefont
  {Bachelard}},\ }\bibfield  {title} {\enquote {\bibinfo {title} {Quantum
  effects in the cooperative scattering of light by atomic clouds},}\ }\href
  {\doibase 10.1103/physreva.95.053625} {\bibfield  {journal} {\bibinfo
  {journal} {Physical Review A}\ }\textbf {\bibinfo {volume} {95}} (\bibinfo
  {year} {2017}),\ 10.1103/physreva.95.053625}\BibitemShut {NoStop}%
\bibitem [{\citenamefont {Scully}\ and\ \citenamefont
  {Zubairy}(1997)}]{Scully1997}%
  \BibitemOpen
  \bibfield  {author} {\bibinfo {author} {\bibfnamefont {M.O.}\ \bibnamefont
  {Scully}}\ and\ \bibinfo {author} {\bibfnamefont {M.S.}\ \bibnamefont
  {Zubairy}},\ }\href {https://books.google.com.br/books?id=20ISsQCKKmQC}
  {\emph {\bibinfo {title} {Quantum Optics}}}\ (\bibinfo  {publisher}
  {Cambridge University Press},\ \bibinfo {year} {1997})\BibitemShut {NoStop}%
\bibitem [{\citenamefont {Labeyrie}\ \emph {et~al.}(2003)\citenamefont
  {Labeyrie}, \citenamefont {Delande}, \citenamefont {M\"{u}ller},
  \citenamefont {Miniatura},\ and\ \citenamefont {Kaiser}}]{Labeyrie2003}%
  \BibitemOpen
  \bibfield  {author} {\bibinfo {author} {\bibfnamefont {Guillaume}\
  \bibnamefont {Labeyrie}}, \bibinfo {author} {\bibfnamefont {Dominique}\
  \bibnamefont {Delande}}, \bibinfo {author} {\bibfnamefont {Cord~A.}\
  \bibnamefont {M\"{u}ller}}, \bibinfo {author} {\bibfnamefont {Christian}\
  \bibnamefont {Miniatura}}, \ and\ \bibinfo {author} {\bibfnamefont {Robin}\
  \bibnamefont {Kaiser}},\ }\bibfield  {title} {\enquote {\bibinfo {title}
  {Coherent backscattering of light by an inhomogeneous cloud of cold atoms},}\
  }\href {\doibase 10.1103/physreva.67.033814} {\bibfield  {journal} {\bibinfo
  {journal} {Physical Review A}\ }\textbf {\bibinfo {volume} {67}} (\bibinfo
  {year} {2003}),\ 10.1103/physreva.67.033814}\BibitemShut {NoStop}%
\end{thebibliography}%

\end{document}